\documentclass[twocolumn,preprintnumbers,pre, superscriptaddress]{revtex4-2}
\usepackage{graphicx, subfigure}
\usepackage{amssymb, amsmath,amssymb,amsfonts}
\usepackage{amsthm,mathrsfs,amsopn}
\usepackage{dcolumn}
\usepackage{bm}
\usepackage{enumerate}
\usepackage[utf8]{inputenc}
\usepackage{mathtools}
\usepackage{tikz}
\graphicspath{{./Figures}}
\theoremstyle{plain} 

\usepackage{tikz}
\usepackage{graphicx}
\usepackage{float}
\usepackage{orcidlink}
\usepackage{xcolor}
\usepackage{ulem}

\begin{document}
\title{Energy–dynamics interplay in temporal networks triggers explosive synchronization}

\author{Romuald Mbonwouo\,\orcidlink{0009-0003-9519-5639}}
\affiliation{Research Unit Condensed Matter, Electronics and Signal Processing,
University of Dschang, P.O. Box 67 Dschang, Cameroon}
\affiliation{MoCLiS Research Group, Dschang, Cameroon}

\author{Steve J. Kongni\,\orcidlink{0000-0001-6036-8905}}
\affiliation{Centre for Audio, Acoustics and Vibration, Faculty of Engineering and IT, University of Technology Sydney, Ultimo NSW 2007, Australia}
\affiliation{MoCLiS Research Group, Dschang, Cameroon.}

\author{Sishu Shankar Muni\,\orcidlink{0000-0001-9545-8345}} 
\affiliation{School of Digital Sciences, Digital University Kerala, Technopark phase-IV
Campus, Mangalapuram, 695317, Thiruvananthapuram, India}

\author{Carmel T. Lambu\,\orcidlink{0009-0009-4269-3929}}
\affiliation{Research Unit Condensed Matter, Electronics and Signal Processing,
University of Dschang, P.O. Box 67 Dschang, Cameroon}
\affiliation{MoCLiS Research Group, Dschang, Cameroon}

\author{Venceslas Nguefoue}
\affiliation{Research Unit Condensed Matter, Electronics and Signal Processing,
University of Dschang, P.O. Box 67 Dschang, Cameroon}
\affiliation{MoCLiS Research Group, Dschang, Cameroon}

\author{Patrick Louodop\,\orcidlink{0000-0002-2975-2420}}
\affiliation{Research Unit Condensed Matter, Electronics and Signal Processing,
University of Dschang, P.O. Box 67 Dschang, Cameroon}
\affiliation{ICTP South American Institute for Fundamental Research, S\~{a}o Paulo State University (UNESP), Instituto de F\'{i}sica T\'{e}orica,Rua Dr. Bento Teobaldo Ferraz 271, Bloco II, Barra Funda, 01140-070 S\~{a}o Paulo, Brazil.}
\affiliation{MoCLiS Research Group, Dschang, Cameroon}

\author{Thierry Njougouo\,\orcidlink{0000-0001-7706-7674}}
\email{thierry.njougouo@imtlucca.it}
\affiliation{IMT School for Advanced Studies Lucca, Italy}
\affiliation{MoCLiS Research Group, Dschang, Cameroon}

\begin{abstract}
Synchronization in networks of coupled oscillators is a fundamental problem in the study of collective behavior. In this paper, we investigate the synchronization transition in networks of coupled dynamical systems from an energetic perspective. Interactions between systems/oscillators are assumed to be governed by one of the following mechanisms: (i) the intrinsic energy $\mathbf{H}$, describing the conservative internal dynamics of isolated systems, and (ii) the dissipative energy $\dot{\mathbf{H}}$, accounting for energy losses and exchanges due to interactions and damping. An energetic threshold is introduced to modulate the network connectivity, so that the topology evolves in time according to the instantaneous energetic similarity between systems, allowing us to analyze how the balance between intrinsic and dissipative energy shapes the transition to synchronization. Using the Rössler and Lorenz systems as representative examples, while keeping the framework general and applicable to other dynamical systems, we explore three representative dynamical regimes: periodic, multiperiodic, and chaotic. This reveals that, the nature of the synchronization transition strongly depends on the interplay between microscopic dynamics and the mesoscopic connectivity structure. In particular, chaotic oscillators coupled through intrinsic energy favor explosive synchronization, corresponding to a first-order transition, whereas periodic and multiperiodic dynamics lead to smooth second-order transitions. In contrast, dissipative-energy-based connectivity suppresses first-order transitions in chaotic networks but can induce second-order transition in multiperiodic systems.
These findings show that explosive synchronization is not solely topology-driven, as often emphasized in scale-free networks, but can also arise from an energy--dynamics--topology feedback controlled by local dynamical complexity and energy-based link activation.
This work provides new insights into how internal oscillator dynamics and coupling mechanisms jointly shape collective organization and dynamical transition patterns in complex systems.
\end{abstract}

\maketitle

\section{Introduction}

Networks of coupled dynamical systems provide a powerful framework for describing collective behaviors in many real-world scenarios such as brain functions \cite{LeVanQuyen2003}, social networks \cite{Friemel2011},  opinion dynamics \cite{Gabbay2007}, electrical networks \cite{Oliver2014}, financial markets \cite{HSIEH1991}, food webs \cite{PahlWostl1997}, climatic networks \cite{Carpi2012}, transportation networks \cite{Pagliara2019}, or gene regulatory networks \cite{Rand2021}. In many of these systems, the collective behavior observed at the macroscopic level cannot be understood solely from the microscopic properties of individual units, but emerges from the interplay between local dynamics and the structure of interactions. Synchronization is one of the most important collective phenomena observed in coupled dynamical systems, because it describes the spontaneous emergence of coherence from initially incoherent units.

In many real-world systems, synchronization is often accompanied by energy exchange, dissipation, or transport between coupled systems or components. Examples include coupled mechanical and physical systems, which are typically dissipative systems  and in which a balance between energy injection and energy loss determines the observed dynamics \cite{GarcaSandoval2015}.
More generally, the relationship between macroscopic behavior, mainly the synchronization and energy flow has attracted increasing attention in nonlinear dynamics, statistical physics, thermodynamics, and network science. In this direction, studies in stochastic energetics have provided a bridge between stochastic dynamical processes and thermodynamic concepts \cite{Sekimoto2010}, while thermodynamic descriptions of collective phase transitions have been developed in globally coupled oscillator and Kuramoto-type systems \cite{Sasa2015}. On the other hand, dynamical phase transitions between synchronization and desynchronization have also been investigated from the perspective of oscillator-network thermodynamics \cite{Imparato2015}. These results indicate that energy may not be considered as a passive quantity associated with the motion of individual units, but can also play an active role in shaping collective behaviors observed at the macroscopic scale.

In biological networks, energy availability plays a fundamental role in their functioning, particularly in neuronal systems where synaptic transmission requires substantial metabolic support. Studies have shown that, neuronal activity is closely coupled to cellular energy metabolism through interactions between neurons and astrocytes, which regulate the supply and utilization of energy substrates such as glucose and lactate. The astrocyte–neuron lactate shuttle mechanism has been proposed as a key pathway supporting neuronal energy demands during synaptic activity \cite{pellerin1998evidence}. Therefore, disruptions in this bioenergetic coupling can significantly alter neural communication and network stability. In particular, abnormalities in metabolic pathways and mitochondrial function have been reported in schizophrenia, suggesting that impaired energy regulation may contribute to synaptic dysfunction and altered brain network dynamics \cite{sullivan2018defects}. These few examples support the claim that, energy availability and dissipation may directly influence the emergence or suppression of coherent collective behaviors.

Several researches have addressed the question of the connection between energy transport, information exchange, and synchronization. For instance, Nicosia et al. studied the interplay between energy transport and synchronization in multilayer networks, with relevance to spontaneous synchronization in the brain \cite{Nicosia2017}. Information exchange and communicability in complex networks have been explored in \cite{West2008,Estrada2012}. 
Although these analyses highlight the relevance of energy in collective dynamics, the way in which local energetic states may dynamically generate the network topology and control the nature of synchronization transitions remains less understood.

According to the literature, a particularly striking form of synchronization is explosive synchronization, where a network of coupled oscillators undergoes an abrupt, first-order-like transition from an incoherent state to a coherent synchronized state and vice versa \cite{Vlasov2015,Skardal2014}. Such abrupt transitions are relevant because of its theoretical framework to understand sudden collective changes in complex systems. Different mechanisms have been proposed to explain explosive synchronization in complex networks, including quenched disorder in oscillator frequencies \cite{Kim2022}, time-delayed coupling in scale-free networks \cite{Peron2012}, bistability in neural networks \cite{Boaretto2019}, changes in basins of attraction \cite{Zou2014}, coexistence between synchronized and unsynchronized states \cite{Leyva2013}, community organization \cite{Lotfi2018}, and adaptive network mechanisms \cite{AvalosGaytn2018}. Although these contributions have significantly enhanced the understanding of explosive synchronization, most existing approaches focus on structural features, imposed correlations, delayed interactions, or prescribed adaptive rules. However, the possibility that explosive synchronization may emerge from an energy-driven temporal reorganization of the network remains comparatively unexplored.

Time-varying networks provide a natural setting for addressing this issue, because many real systems evolve not only through their node dynamics, but also through changes in their interaction topology and coupling strengths \cite{Perra2012,ghosh2022synchronized,petit2017theory,kohar2014synchronization,njougouo2020effects}. In such systems, connections or links between systems may appear, disappear, or change over time, so the relevant question is not only \textit{who interacts with whom}, but also \textit{when} and \textit{under which dynamical conditions} interactions occur. Synchronization in time-varying networks has been investigated in several contexts, including FitzHugh--Nagumo neuronal networks under external effects \cite{Huang2025}, ring networks of neurons with additive noise and space--time-dependent topology \cite{Ghosh2023}. Such studies highlight how structural and dynamical fluctuations can influence collective neuronal behavior. Similarly, biological synchronization may emerge in response to temporally structured stimuli, such as eye-blinking synchronization with musical beats \cite{Wu2025}. Nevertheless, a general understanding of how energy-based temporal connectivity influences explosive synchronization is still very less explored.

The originality of the present work is to introduce an energy-driven temporal-network framework for studying the transition to synchronization. In contrast to the classical approaches reported in the literature, where the network topology is prescribed a priori and remains static, here the links between systems are dynamically activated according to energetic similarities between nodes. The network topology is therefore not an external constraint, but an emergent feature determined by the instantaneous energetic state of the systems. This approach provides a physically interpretable mechanism in which effective connections emerge between nodes whose intrinsic energies or dissipative activities differ by less than a defined threshold. As a result, local energetic compatibility determines the mesoscopic temporal organization of the network and, ultimately, controls the nature of the macroscopic synchronization transition. To build this physically interpretable framework, we use a Hamiltonian decomposition of each system of the network into conservative and dissipative parts. This naturally defines two energy-based rules for temporal link formation: one based on intrinsic energy and the other on dissipative energy. This allows us to compare two routes to synchronization, respectively governed by compatibility in the internal energetic state of the oscillators and by compatibility in their dissipative activity. The application of this formalism to the Rössler and Lorenz systems, shows that intrinsic-energy-based connectivity favors explosive synchronization in chaotic regimes. In contrast, dissipative-energy-based connectivity tends to suppress this first order transition in chaotic systems but may still induce abrupt-like transitions in multiperiodic regimes when the network becomes densely connected.

The paper is organized as follows. Section~\ref{sec:MDM} introduces the model within an energy-based framework, where the network topology adapts according to similarities in intrinsic or dissipative energy among oscillators. In Section~\ref{sec:sec3}, we analyze the role of this energy-based interactions in the synchronization transition. Section~\ref{sec:sec4} examines the influence of the internal state of the oscillators on the transition to synchronization, while Sec.~\ref{sec::LO} extends the analysis to a network of Lorenz system. Finally, Section~\ref{sec: concl} presents the conclusion.

\section{Model description and methodology}\label{sec:MDM}

This study explores the phenomenon of synchronization within networks composed of nonlinear and chaotic systems, through the lens of energy relationships, specifically intrinsic energy and dissipative energy. Intrinsic energy characterizes the internal dynamics of an isolated system, whereas dissipative energy accounts for energy losses and exchanges arising from interaction and damping mechanisms~\cite{sarasola2004energy,ginoux2025energy}.
In a network, coupled systems with similar intrinsic energies enhances energy exchange efficiency through spectral proximity, promoting resonance and facilitating synchronization by minimizing frequency mismatches, a mechanism well documented in nonlinear oscillator theory and phase synchronization models such as Kuramoto’s framework~\cite{kuramoto2003chemical,pikovsky2001universal,kongni2023phase,kongni2024expected}. Examples are well known in neuroscience, where neurons with close intrinsic firing frequencies synchronize more readily, supporting coherent rhythms associated with cognition and memory~\cite{buzsaki2006rhythms}; in power grids, generators with similar natural frequencies improve  stability and reduce the risk of cascading failures~\cite{motter2013spontaneous}. In the same vein, coupling systems with comparable dissipative energies which could mean similar damping rates or relaxation times, ensures compatible temporal scales, allowing sustained interactions without rapid dominance or suppression, which is crucial in biological networks where matched dissipation supports stable metabolic or cellular coordination~\cite{alon2019introduction}. In complex networks, such energetic compatibility could enhance collective stability and cluster formation.
The primary objective is to understand how energy exchange can either promote or hinder interactions between systems, thereby shaping collective behaviors within the entire network.

\subsection{Hamiltonian formalism}\label{sec::HF}

Let us consider a temporal network, i.e., a network whose connectivity may evolve over time, composed of $N$ nodes, each hosting an identical dynamical system. The dynamics of node $j$ is described, in its general form, by Eq.~\ref{eq::eq1}:

\begin{equation}
\frac{d\mathbf{X}_j(t)}{dt} = \mathbf{F}(\mathbf{X}_j(t)) + \epsilon \sum_{i=1}^N \mathbf{G}_{ji}(t) \left(\mathbf{W}(\mathbf{X}_i) - \mathbf{W}(\mathbf{X}_j) \right).
\label{eq::eq1}
\end{equation}
Where $j=1, \cdots, N$, $\mathbf{X}_j \in \Re^m $ represents the state vector of the $j$-th system, $\mathbf{F}(\mathbf{X}_j): \Re^m \to \Re^m $ denotes the local dynamics of each system, $\mathbf{W}(\cdot): \Re^m \to \Re^m $ represents the interaction or the coupling function between systems $j$ and $i$,  $\mathbf{G}: \Re^{N \times N}$ is the adjacency matrix encoding the network structure, and $\epsilon$ represents the coupling strength. The adjacency matrix $\mathbf{G}$ is constructed based on the energy function developed below.

Recent works such as refs.~\cite{jiang2024hamilton,sarasola2004energy,ginoux2025energy,njougouo2020dynamics} propose methods based on Helmholtz's theorem, which allow the derivation of an energy function for any $m$-dimensional autonomous dynamical system. According to Helmholtz's theorem, any local $m$-dimensional autonomous system, expressed as
\[
\frac{d\mathbf{X}_j(t)}{dt} = \mathbf{F}(\mathbf{X}_j(t)),
\]
can be decomposed into the sum of two components (see Eq.~\ref{eq::eq2}):\\
\begin{equation}
\textcolor{black}{\mathbf{F}(\mathbf{X}_j(t)) = \mathbf{F}_c(\mathbf{X}_j(t)) + \mathbf{F}_d(\mathbf{X}_j(t)),}
\label{eq::eq2}
\end{equation}
i) a divergence-free vector field $\mathbf{F}_c(\mathbf{X}_j(t))$, often referred to as the conservative component, which captures the entire rotational part of $\mathbf{F}(\mathbf{X}_j(t))$;\\
ii) a gradient vector field $\mathbf{F}_d(\mathbf{X}_j(t))$, commonly called the dissipative component, which accounts for its entire divergence.

According to this decomposition, the local dynamic of any system can be written in a following generalized Hamiltonian form as:
\begin{equation}
\frac{d\mathbf{X}_j(t)}{dt}  = \left[ \mathbf{J}(\mathbf{X}_j(t)) + \mathbf{R} (\mathbf{X}_j(t))\right]\nabla  \mathbf{H}.
\label{eq::eqHF}
\end{equation}
where $\nabla \mathbf{H}$ is the gradient vector of a smooth energy function $\mathbf{H}(\mathbf{X}_j)$, $\mathbf{J}(\mathbf{X}_j)$ is a skew-symmetric matrix and $\mathbf{R}(\mathbf{X}_j)$ is a symmetric matrix associated with the node $j$. Therefore, the Hamiltonian energy function can be expressed by Eq.~\ref{eq::eh}:
\begin{equation}
    \frac{d\mathbf{H}}{dt} = \nabla\mathbf{H}^\top \mathbf{R}(\mathbf{X}_j) \nabla \mathbf{H},
\quad
\nabla \mathbf{H}^\top \mathbf{J}(\mathbf{X}_j) \nabla \mathbf{H} = 0.
\label{eq::eh}
\end{equation}
The decomposition of the vector field given by Eq.~\ref{eq::eq2} into its gradient and rotational components allows the corresponding energy function to be explicitly determined as:

\begin{equation}
\nabla  \mathbf{H}^T \mathbf{F}_c(\mathbf{X}_j(t)) = 0.
\label{eq::eq3}
\end{equation}
This corresponds to the energy associated with the divergence-free vector field 
$\mathbf{F}_c(\mathbf{X}_j(t))$, representing a conserved quantity that remains constant along the trajectories of the system. It reflects the intrinsic rotational properties of the vector field and is conserved due to its divergence-free nature.
\begin{equation}
\textcolor{black}{\dot{\mathbf{H}}(\mathbf{X}_j) =  \nabla \mathbf{H}^T \mathbf{F}_d(\mathbf{X}_j(t)).}
\label{eq::eq4}
\end{equation}
Eq.~\ref{eq::eq4} express the time derivative of the energy, which represents the energy dissipated—either actively or passively—through the gradient vector field $\mathbf{F}_d(\mathbf{X}_j(t))$. 
This term can be understood as the instantaneous rate of work performed by the energy gradient, quantifying energy transfer, redistribution, or loss due to the divergence of the vector field. 
It thus provides a quantitative measure of the system's dynamic response to changes in its energetic state.

\subsection{Energy  and network structure}

To relate the energy dynamics of each node to the network structure, we analyse how the energies computed using the Hamiltonian formalism described above can inform the formation of connections, i.e., the adjacency matrix $\mathbf{G}(t)$ of the network at each time step $t$. 
In the following, we consider two hypotheses for constructing the time-varying adjacency matrix $\mathbf{G}(t)$. Specifically, the intrinsic energy $\mathbf{H}_j(t)$ characterizes the instantaneous internal state (or local dynamics) of node $j$, whereas the dissipative (or interaction) contribution is reflected by the temporal evolution of this energy, which we capture through the rate $\dot{\mathbf{H}}_j(t)$. For the sake of notational simplicity, we omit the explicit dependence of the energy functions on the state variables. Thus, instead of writing $\mathbf{H}(\mathbf{X}_j)$(resp. $\dot{\mathbf{H}}(\mathbf{X}_j)$) for each node $j$, we simply denote them by $\mathbf{H}_j$(resp. $\dot{\mathbf{H}}_j$).
\\
i) Under the intrinsic-energy hypothesis, two nodes $j$ and $i$ are connected at time $t$ if their intrinsic energies are sufficiently close:
\begin{equation}
\label{eq::eq5} 
\mathbf{G}_{ji}(t) = 
    \begin{dcases}
        1, & \text{if} \quad \lvert \mathbf{H}_j(t) - \mathbf{H}_i(t) \rvert < \delta, \\
        0, & \text{otherwise}.
    \end{dcases}
\end{equation}
ii) However, under the dissipative (interaction) hypothesis, connectivity is instead determined by similarity in the energy rates, i.e., comparable dissipative/interaction activity:
\begin{equation}
\label{eq::eq6} 
\mathbf{G}_{ji}(t) = 
    \begin{dcases}
        1, & \text{if} \quad \lvert \dot{\mathbf{H}}_j(t) - \dot{\mathbf{H}}_i(t) \rvert < \delta_D, \\
        0, & \text{otherwise}.
    \end{dcases}
\end{equation}

In the energetic approach of Sarasola et al.~\cite{sarasola2004energy}, the isosurfaces $\mathbf{H} = \mathbf{constant}$ represent energy levels in phase space: the conservative flow remains confined to a given isosurface, whereas dissipation induces crossings between isosurfaces through the time evolution of $\mathbf{H}$ (via $\mathbf{\dot H}$)~ \cite{sarasola2004energy}.
Accordingly, the thresholds $\delta$ and $\delta_D$ can thus be interpreted as \textit{thickness tolerances} that define, respectively, an energy band around a level set and a tolerance on the similarity of dissipative activity. 
Therefore, a small $\delta$ restricts connectivity to nodes lying at nearly the same energy level, while a larger $\delta$ merges several levels and blurs the interpretation in terms of a common isosurface. 
Likewise, $\delta_D$ sets the resolution over \textit{iso-rate} regimes ($\mathbf{\dot H}=\mathbf{cste}$), grouping nodes that cross energy isosurfaces with comparable intensity. Consequently, the choice of the values of $\delta$ and $\delta_D$ requires prior knowledge (or empirical estimates) of the typical amplitudes and variability ranges of $\mathbf{H}$ and $\mathbf{\dot H}$, so that these tolerances are chosen consistently with the energy scales of the system.

This approach implies that the network evolves dynamically based on the energy states of the systems, reflecting how changes in energy influence the connectivity and interactions within the network. This method allows us to model and analyze how the structure of the network adapts in response to the energetic properties of its constituent systems, thereby providing insights into the interplay between energy dynamics and network structure.

\subsection{Dynamics of a single node: R\"ossler oscillator}
\begin{figure}[htp!]
    \centering
    \begin{tabular}{cc}
    \includegraphics[width=0.228\textwidth]{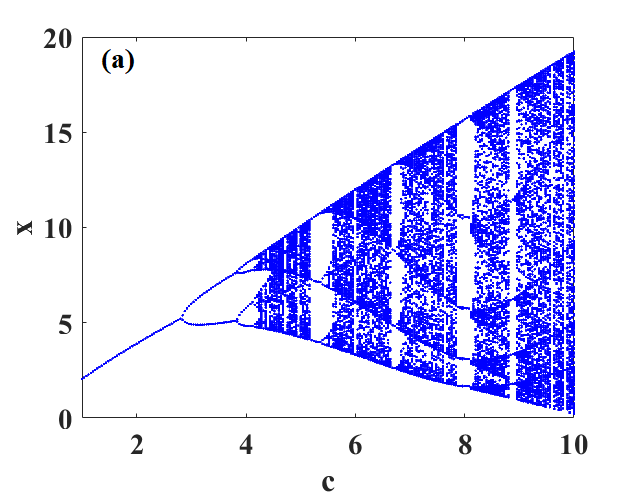}&
    \includegraphics[width=0.24\textwidth]{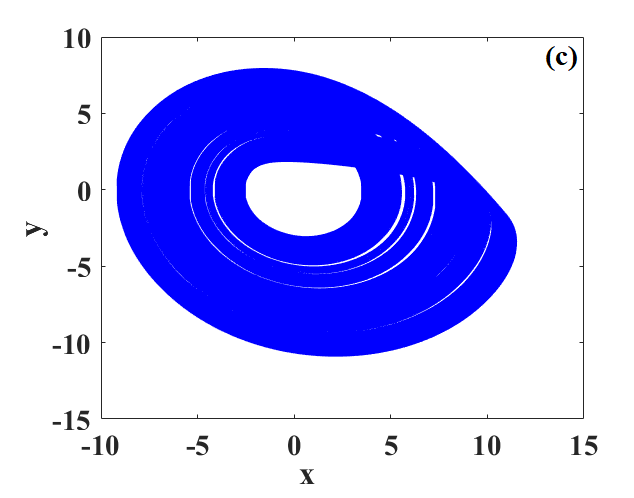}\\
    \includegraphics[width=0.248\textwidth]{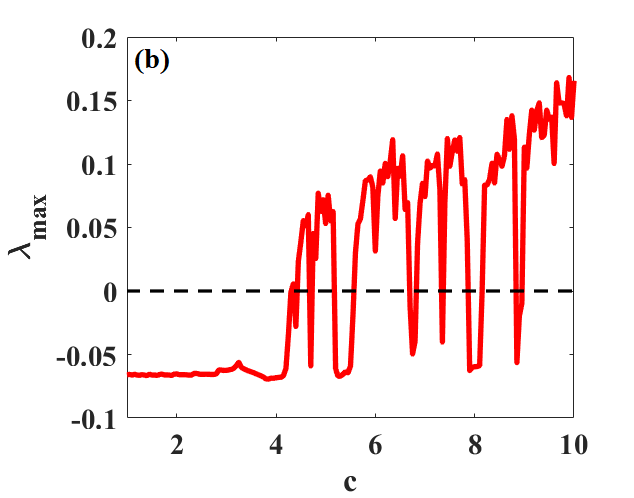} &
     \includegraphics[width=0.24\textwidth]{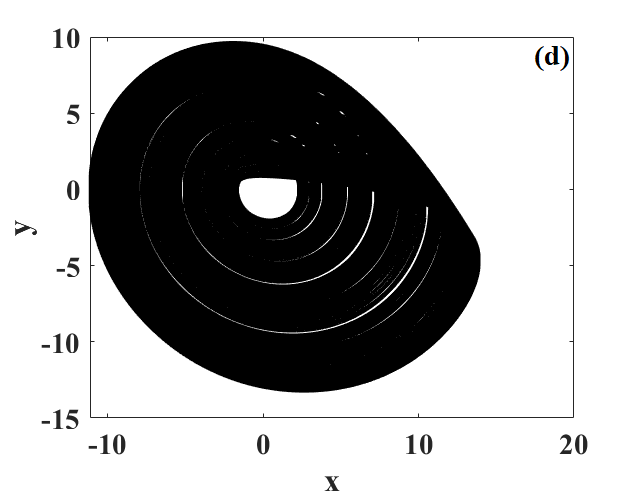} 
    \end{tabular}
    \caption{Illustration of the dynamical behavior of the Rössler system as a function of the parameter $c$. (a) Bifurcation diagram and (b) Maximum Lyapunov exponent with $a = 0.2$ and  $b = 0.2$. (c) and (d) show Rössler attractors for $c = 5.7$ (blue) and $c = 7$ (black), respectively.}
    \label{fig1}
\end{figure}

In the absence of interactions, i.e., coupling, the evolution of each node, as introduced in the general framework of Eq.~\ref{eq::eq1}, reduces to the three-dimensional R\"ossler oscillator~\cite{ROSSLER1976397}, whose equations are given in Eq.~\ref{eq::eq1r}.

\begin{equation}
    \begin{cases}
        \dot{x} = -y - z \\
        \dot{y} = x + a y \\
        \dot{z} = b + z(x - c)
    \end{cases}
    \label{eq::eq1r}
\end{equation}
where $x$, $y$, and $z$ represent the state variables, while $a = 0.2$, $b = 0.2$, and $c = 5.7$ are characteristic parameters of the system. For these parameter values, the system exhibits chaotic behavior \cite{ROSSLER1976397}. To illustrate how the dynamics of the isolated oscillator depend on the parameter $c$, we show in Fig.~\ref{fig1} its behavior for $a = b = 0.2$.
Fig.~\ref{fig1}(a) shows the bifurcation diagram, where various dynamical regimes such as periodicity, multiperiodicity, and chaos can be observed. These different behaviors are further supported by the plot of the maximal Lyapunov exponent $\lambda_{\max}$  defined by Eq.~\ref{eq::Lam} an shown in Fig.~\ref{fig1}(b):
\begin{equation}
\lambda_{\max} = \lim_{t \to \infty} \lim_{\| \delta \mathbf{X}(0) \| \to 0} \frac{1}{t} \ln \frac{\| \delta \mathbf{X}(t) \|}{\| \delta \mathbf{X}(0) \|}.
\label{eq::Lam}
\end{equation}
Periodic and multiperiodic states are characterized by $\lambda_{\max} < 0$, whereas chaotic states are identified by $\lambda_{\max} > 0$, leading to a dense set of points in the bifurcation diagram.
To further illustrate these findings, Fig.~\ref{fig1}(c) presents the phase portrait for $c = 5.7$—a classical parameter value known to produce chaos in the R\"ossler system—showing $\lambda_{\max} > 0$. Fig.~\ref{fig1}(d) displays the corresponding attractor for $c = 7$, which also yields $\lambda_{\max} > 0$. In both cases, the system remains chaotic, corroborating the predictions of the bifurcation diagram and the Lyapunov exponent analysis.

\section{interplay between energy and synchronization} \label{sec:sec3}
Let us consider a network of $N$ dynamical systems, where the dynamics of each node evolves according to the general framework of Eq.~\ref{eq::eq1}, with the local dynamics given by the R\"ossler system and coupled through the interaction term. The overall dynamics of the network can be captured by Eq.~\ref{eq::eq9}, which incorporates both the local R\"ossler dynamics and the interactions among the nodes.

\begin{equation}
\label{eq::eq9} 
    \begin{dcases}
        \dot x_j = - y_j - z_j + \epsilon \sum_{i=1}^N G_{ji}(t) \left(x_i - x_j\right)\\
        \dot y_j = x_j + a y_j  \\  
        \dot z_j = b + (x_j - c)z_j.
    \end{dcases}\,
\end{equation}
The connection between two oscillators, $j$ and $i$, at time $t$  encoded in the matrix $G(t)$ is defined either by their intrinsic energy (see Eq.~\ref{eq::eq5}), which characterizes their local dynamics, or by their dissipative energy (see Eq.~\ref{eq::eq6}), which describes the coupling-induced exchanges and losses. From a physical perspective, coupling systems with similar intrinsic energies enhances resonance and synchronization by reducing frequency mismatches and improving energy transfer efficiency~\cite{kuramoto2003chemical,pikovsky2001universal,kongni2023phase,kongni2024expected}. Similarly, coupling systems with similar dissipative energies, associated with close damping rates or relaxation times, ensures compatible temporal scales and sustained interactions.

\subsection{Interactions driven by intrinsic energy}\label{sec::secIE} 

We consider here that the coupling between the elements of the network is determined by their intrinsic energy. In other words, the network structure encoded by the adjacency matrix $\mathbf{G}(t)$ at each time $t$ is generated according to Eq.~\ref{eq::eq5}, where $\mathbf{H}_j(t)$ denotes the intrinsic energy of system $j$ at time $t$. To derive the analytical expression of the intrinsic energy $\mathbf{H}_j$ for each system $j$ in the network, we follow the method proposed by Sarasola et al.~\cite{sarasola2004energy}, as detailed in Sec.~\ref{sec::HF}. According to Eq.~\ref{eq::eq2}, the local dynamics of each R\"ossler system in the network at each time step $t$ can be decomposed into the sum of two vector fields: a divergence-free vector field $\mathbf{F}_c(\mathbf{X}_j)$ and a gradient vector field $\mathbf{F}_d(\mathbf{X}_j)$, commonly referred to as the dissipative component, which is expressed in Eq.~\ref{eqfd}.
\begin{equation}
\begin{aligned}
\mathbf{F}_c(\mathbf{X}_j) &=
\begin{pmatrix}
-y_j - z_j - \tfrac{1}{2} z_j^2 \\
x_j \\
b
\end{pmatrix}, \quad \text{and}
\\[6pt]
\mathbf{F}_d(\mathbf{X}_j) &=
\begin{pmatrix}
\tfrac{1}{2} z_j^2 \\
a y_j \\
(x_j - c) z_j
\end{pmatrix},
\end{aligned}
\label{eqfd}
\end{equation}
where $\mathbf{X}_j = (x_j, y_j, z_j)$ denotes the state vector of oscillator $j$.

Substituting $\mathbf{F}_c(\mathbf{X}_j)$ into Eq.~\ref{eq::eq3}, we obtain the following partial differential equation, given in Eq.~\ref{eq::eq10}:

\begin{equation}
-(y_j + z_j + \frac{1}{2} z_j^2) \frac{\partial \mathbf{H}_j}{\partial x_j} + x_j \frac{\partial \mathbf{H}_j}{\partial y_j} + b \frac{\partial \mathbf{H}_j}{\partial z_j} = 0.
\label{eq::eq10}
\end{equation}
By solving Eq.~\ref{eq::eq10}, we obtain the analytical expression of the intrinsic energy for each node $j$, as given in Eq.~\ref{eq::eq11}.
\begin{equation}
\mathbf{H}_j = \frac{1}{2} \left[ \left(x_j + b(z_j + 1)\right)^2 + \left( y_j + \frac{1}{2}z_j^2 + z_j - b^2 \right)^2 \right].
\label{eq::eq11}
\end{equation}
According to Eq.~\ref{eq::eq5}, gradually increasing the threshold parameter $\delta$ enables a smooth transition from a sparse network (for $\delta \to 0$) to a fully connected network (for $\delta \to +\infty$). This $\delta$ defines the tolerance in intrinsic energy differences required for two nodes to interact: a small $\delta$ allows connections only between nodes with very similar energies, while a large $\delta$ allows connection of nodes with increasingly different energies. 
In the following, we consider a network of $100$ R\"ossler oscillators, where the structure $\mathbf{G}(t)$ is modulated by the intrinsic energy of the systems composing the network. The initial conditions of the oscillators are randomly sampled within the interval $[-1, 1]$. The system dynamics are solved numerically using the fourth-order Runge–Kutta algorithm.

The parameters used for the Rössler systems are $a = 0.2$, $b = 0.2$, and $c = 7$.
To illustrate the transition to phase synchronization in the network, we use the order parameter, a concept originally introduced by Kuramoto and Battogtokh~\cite{kuramoto2002coexistence}, which provides a powerful measure for characterizing phase coherence in systems of coupled oscillators. Its computation relies on extracting the instantaneous phase from each individual time series within the system. 
To determine the phase of a given time-dependent signal $s(\tau)$, we construct its associated analytic signal using the Hilbert transform $\tilde{s}(\tau)$. The analytic signal is defined as:
\begin{equation*}\label{ht}
\psi(\tau) = s(\tau) + i\,\tilde{s}(\tau) = r(\tau) e^{i\varphi(\tau)},
\end{equation*}
where $i^2 = -1$, $r(\tau)$ denotes the instantaneous amplitude, and $\varphi(\tau)$ the instantaneous phase of the signal $s(\tau)$. 
For each oscillator $j$, the instantaneous phase $\varphi_j(\tau)$ is thus given by:
\begin{equation*}\label{pha}
\varphi_j(\tau) = \tan^{-1}\left(\frac{\tilde{s}_j(\tau)}{s_j(\tau)}\right).
\end{equation*}
The global phase coherence of a network with $N$ oscillators is then captured by the order parameter $r$ given by:
\begin{equation}\label{op}
r = \left| \frac{1}{N} \sum_{j=1}^N e^{i\varphi_j} \right|.
\end{equation}

\begin{figure*}[htp!]
    \centering
    \begin{tabular}{ccc}
    \includegraphics[width=0.325\textwidth]{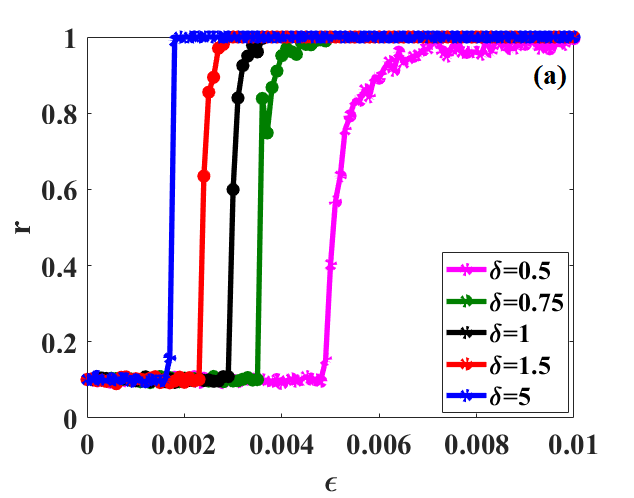}&
    \includegraphics[width=0.325\textwidth]{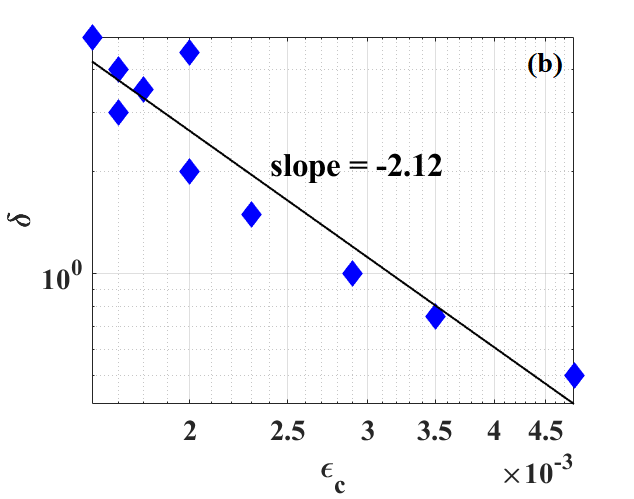}&
    \includegraphics[width=0.325\textwidth]{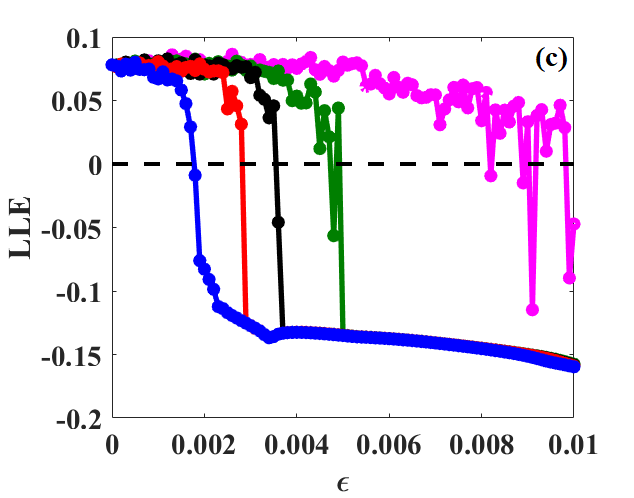}
    \end{tabular}
    \caption{Transition to phase synchronization and dynamical stability analysis : (a) The evolution of the order parameter $r$ as a function of the coupling strength $\epsilon$, (b) Semi–log representation of the distribution of critical coupling strengths $\epsilon_c$ obtained for different realizations of the threshold parameter $\delta$, revealing a power-law behavior, and (c) Largest Lyapunov exponent as a function of the coupling strength $\epsilon$, characterizing the stability of the synchronized states and the underlying dynamical regimes.}
    \label{fig2}
\end{figure*}

To better emphasize the impact of intrinsic energy in shaping the collective dynamics, Fig.~\ref{fig2} illustrates the transition to synchronization in this chaotic oscillator network for different values of the threshold parameter $\delta$ (which modulates the network connectivity):
\begin{equation*}
    \delta \in \{0.5,\, 0.75,\, 1,\, 1.5,\, 5\},
\end{equation*}
as a function of the coupling strength $\epsilon$.
In Fig.~\ref{fig2}(a, c), each colored curve corresponds to a specific value of $\delta$, according to the following color scheme: magenta for $\delta = 0.5$, green for $\delta = 0.75$, black for $\delta = 1$, red for $\delta = 1.5$, and blue for $\delta = 5$. 
This synchronization analysis reveals the existence of a critical value $\epsilon_c$ of the coupling strength $\epsilon$ for each value of the threshold $\delta$. This critical value corresponds to a change in the collective behavior, namely the transition of the order parameter $r$ from $r \approx 0$, associated with decoherence or desynchronization, to $r \to 1$, indicating the emergence of a globally coherent state. It also appears that the nature of this transition strongly depends on $\delta$. When $\delta$ is small, the connectivity is selective: only systems with very close intrinsic energies can interact, resulting in a sparser network structure. Synchronization then emerges progressively, and a stronger coupling is required to achieve global synchronization through the gradual buildup of local correlations between nearly resonant elements, leading to a continuous, second-order-like transition. As $\delta$ increases, the energetic constraint is relaxed and systems with more distinct energy levels can connect; the network thus becomes more heterogeneous and more densely connected. This increased interaction range enhances collective cooperativity, lowers the critical coupling threshold $\epsilon_c$, and makes the transition increasingly abrupt and discontinuous, characteristic of a first-order transition~\cite{gomez2011explosive, Leyva2012}.

\begin{figure}[htp!]
    \centering
    \begin{tabular}{c}
    \includegraphics[width=0.45\textwidth]{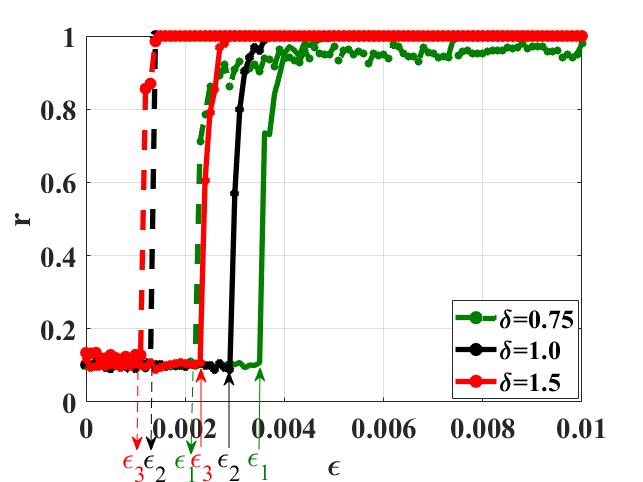}
    \end{tabular}
    \caption{Hysteresis phenomena associated with explosive synchronization transitions as a function of the coupling strength $\epsilon$ for three values of the threshold parameter $\delta \in \{0.75, 1.0, 1.5\}$. Solid curves correspond to the forward (increasing $\epsilon$) process, while dashed curves represent the backward (decreasing $\epsilon$) process.}
    \label{fig2hy}
\end{figure}

Such a first-order transition is often accompanied by a hysteresis effect, as shown in Fig.~\ref{fig2hy} (displayed for three values $\delta \in \{0.75, 1.0, 1.5\}$ for clarity, although hysteresis is also observed for the other $\delta$ values), reflecting the dependence of the system on its history. When $\epsilon$ is gradually increased, synchronization appears at a critical value $\epsilon_c^{\uparrow}$ (see the jump points of the solid curves). Conversely, when $\epsilon$ is decreased, desynchronization occurs at a lower critical value $\epsilon_c^{\downarrow}$ (see the jump points of the dashed curves). The interval $\epsilon_c^{\downarrow} < \epsilon < \epsilon_c^{\uparrow}$ defines a bistable region in which two stable states—an ordered state and a disordered state—can coexist for the same macroscopic parameters.
Thus, this energetic threshold $\delta$ controls both the effective structure of the network and the nature of the order–disorder transition. In Appendix~\ref{Apen}, we present a numerical study illustrating that the explosive character of the synchronization is preserved as the system size increases, thereby supporting its persistence in the thermodynamic limit.

In Fig.~\ref{fig2}(b), the relationship between the critical coupling strength $\epsilon_c$ and the threshold parameter $\delta$ highlights the interdependence between the network connectivity or topology and the onset of synchronization. The observed \textit{power-law} relation, $\delta \propto \epsilon_c^{-\eta}$ with an exponent $\eta \approx 2.12$, indicates that small variations in $\epsilon_c$ can cause large changes in $\delta$. This means that, as the coupling strength required to trigger global synchronization increases, the network must become more selective (smaller $\delta$, i.e., connections occur only between elements with very close intrinsic energies) for the transition to occur. Conversely, when the coupling is sufficiently strong, even networks with larger $\delta$—that is, more heterogeneous connections between elements with different intrinsic energies—can synchronize. This relationship also shows that the effective network structure (via $\delta$) and the interaction strength ($\epsilon_c$) are closely linked in determining the emergence of order. A large exponent $\eta$ implies high sensitivity (as the network becomes sparser): small increases in $\epsilon_c$ drastically reduce the effective connectivity, favoring selective interactions between nearly resonant elements and leading to gradual, second-order-like transitions. Conversely, low values of $\epsilon_c$ correspond to less restrictive connectivity and then densely connected network, allowing for abrupt or explosive, first-order synchronization often accompanied by hysteresis. Finally, this power-law relation presented in Fig.~\ref{fig2}(b) quantifies how microscopic interaction parameters control the macroscopic collective dynamics, i.e., synchronization, showing that both the threshold $\delta$ and the critical coupling $\epsilon_c$ govern \textit{when} synchronization emerges and \textit{how} the system organizes collectively, whether gradually or explosively.

Beyond the phase synchronization phenomena presented in Figs.~\ref{fig2}(a) and \ref{fig2hy}, the model also exhibits complete or full synchronization states, whose stability is analyzed using the Master Stability Function (MSF) developed by Pecora and Carroll~\cite{pecoraPhysRevLett}; the details of this method are provided in Appendix~\ref{Apen1}. The stability of the complete synchronization manifold is assessed via the sign of the Largest Lyapunov Exponent (LLE): a negative LLE indicates stable synchronization, whereas a positive LLE corresponds to instability. To illustrate this, Fig.~\ref{fig2}(c) shows the evolution of the LLE evaluate numerically for four representative values of the threshold parameter $\delta$ used in Fig.~\ref{fig2}(a). The results lead to the same conclusions drawn from the phase synchronization analysis: smaller values of $\delta$ correspond to more selective connectivity, requiring stronger coupling to achieve stable synchronization, whereas larger values of $\delta$ allow the synchronized state to stabilize more easily at lower coupling strengths. An important aspect of this result is that the LLE exhibits a transition behavior very similar to the explosive or abrupt transition observed in the phase synchronization diagrams of Fig.~\ref{fig2}. In particular, for sufficiently large $\delta$, the LLE undergoes an abrupt change from positive to negative values, confirming that the explosive nature of the transition is not limited to phase coherence but also manifests at the level of full dynamical stability.

\begin{figure}[htp!]
    \centering
    \begin{tabular}{c}
    \includegraphics[width=0.5\textwidth]{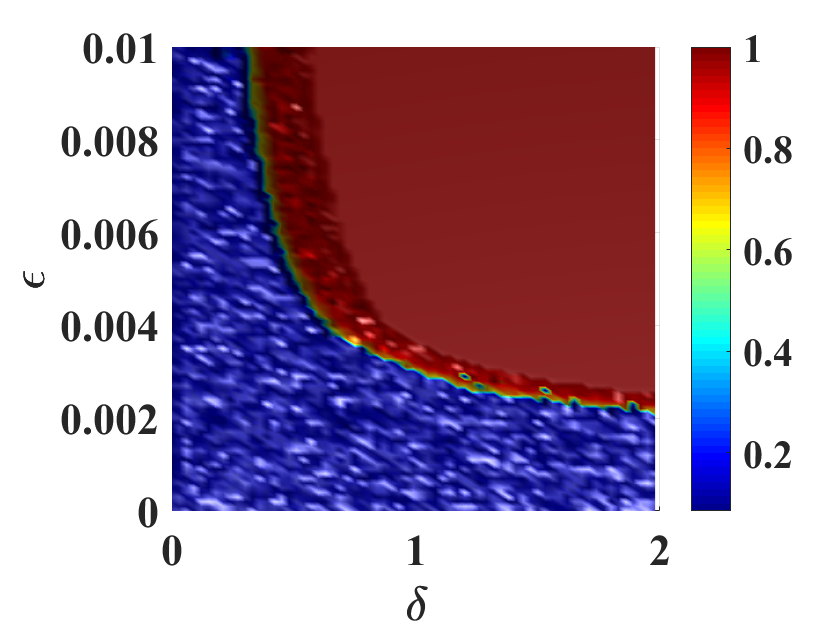}
    \end{tabular}
    \caption{Two-dimensional representation of the order parameter in the $(\delta,\epsilon)$ plane, showing the onset of synchronization and the emergence of explosive transitions. }
    \label{fig2dhmin}
\end{figure}

A more general analysis of the dependence of synchronization on $\delta$, which modulates the network topology, and on $\epsilon$, which controls the coupling strength, is performed to better illustrate their combined impact not only on the emergence of synchronization, but also on the nature of the transition and on the critical threshold $\epsilon_c$. To this end, Fig.~\ref{fig2dhmin} presents the evolution of the order parameter under a double variation of  $\delta$ and $\epsilon$, where the colorbar represents the magnitude of the order parameter $r$, characterizing the level of phase synchronization.

The result clearly reveals two distinct regions: a blue region, where $r$ remains small and corresponds to the desynchronized regime, and a red region, where $r \to 1$, indicating phase synchronization. The boundary between these two regions reflects the synchronization transition line in the $(\delta,\epsilon)$ parameter space. A key observation is that for small values of $\delta$, the critical coupling required to reach synchronization is significantly larger, consistent with a sparse and highly selective network where only oscillators with very close intrinsic energies interact. In this regime, the transition from the blue to the red domain occurs smoothly, through intermediate values of $r$ (even if it is close to one), indicating a gradual, second-order-like transition similar to the case $\delta = 0.5$ in Fig.~\ref{fig2}(a).\\
In contrast, for larger values of $\delta$, the energetic constraint is relaxed and the network becomes more densely and heterogeneously connected. As a result, synchronization is achieved at lower values of $\epsilon$, and the transition from the desynchronized (blue) to the synchronized (red) region becomes much more abrupt. The abrupt change in $r$ across the transition line is the signature of a first-order transition, consistent with the explosive synchronization behavior discussed previously.

\begin{figure}[htp!]
    \centering
    \includegraphics[width=0.35\textwidth]{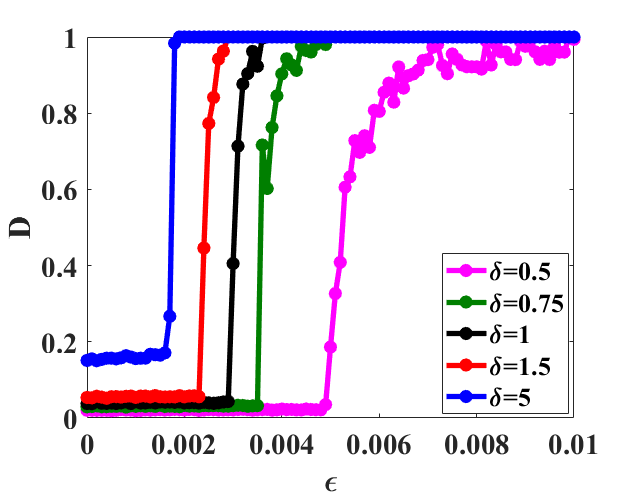}
    \caption{Link density of the network as a function of coupling strength for various values of $\delta$.}
    \label{fig3}
\end{figure}

Let us now investigate the relationship between the collective dynamics on the nodes and the network structure. To this end, Fig.~\ref{fig3} presents the average link density $D$ \cite{bedru2020big} (defined in Eq.~\ref{eq::eqd}) as a function of the coupling strength $\epsilon$ for different values of the connectivity threshold $\delta$.
\begin{equation}\label{eq::eqd}
D = \frac{1}{T} \sum_{t=1}^{T} \left( \frac{1}{N(N-1)} \sum_{j\ne i}^{N} \mathbf{G}_{ji}(t) \right)
\end{equation}

This measure captures how the network structure evolves as a function of the system dynamics. Numerical analysis reveals that the evolution of $D$ with respect to $\epsilon$ closely mirrors the behavior of the order parameter shown in Fig.~\ref{fig2}. This similarity arises because the network topology—specifically the number of effective connections—is directly governed by the oscillators’ energies. When the network is in an incoherent (unsynchronized) state, the oscillators display disparate and typically high energy values, and as a consequence, only a small number of pairs satisfy the connection criterion, leading to a sparse network and very low values of $D$. In contrast, as synchronization progressively emerges, the oscillators’ energies become closer and more homogeneous. This energetic alignment allows more connections to form, producing a rapid increase in network density and thus in $D$, which approaches unity in the fully synchronized regime. Therefore, the monotonic increase of $D$ with $\epsilon$ provides direct evidence that the energy distribution dynamically reshapes the network structure. This result indicates the strong interplay between dynamics and topology: the coupling strength not only drives collective synchronization but also reorganizes the underlying connectivity, reinforcing the transition from a sparse, incoherent regime to a dense, coherent state.

\subsection{Interactions driven by dissipative energy}
In this section, the network structure is governed by the dissipative energy, which corresponds to the energy exchanged through interactions between the oscillators. The network topology is constructed according to the relation defined in Eq.~\ref{eq::eq6}, where $\dot{\mathbf H}$ denotes the dissipative energy of the considered system. To derive the analytical expression of this dissipative energy, we substitute $\mathbf F_d( \mathbf X_j)$, defined in Eq.~\ref{eqfd}, and the Hamiltonian $\mathbf H$, defined in Eq.~\ref{eq::eq10}, into Eq.~\ref{eq::eq6}. This procedure yields the final expression for $\dot{\mathbf H}_j$ for oscillator $j$, as given in Eq.~\ref{eq::eqdis}.

\begin{equation}
\begin{split}
    \dot{\mathbf{H}}_j = \frac{1}{2}z_j^2 + bz_j(x_j - c)x_j + b(z_j+1) + \\
    (ay_j + z_j (x_j-c)(z_j+1))(y_j + \frac{1}{2} z_j^2 + z_j - b^2)
\end{split}
    \label{eq::eqdis}
\end{equation}
An analysis analogous to that performed in Figs.~\ref{fig2}(a,c) for the case of coupling through intrinsic energy is carried out here using the dissipative energy, in order to provide a comparative study between the two coupling mechanisms.

\begin{figure}[htp!]
    \centering
    \begin{tabular}{cc}
       \includegraphics[width=0.25\textwidth]{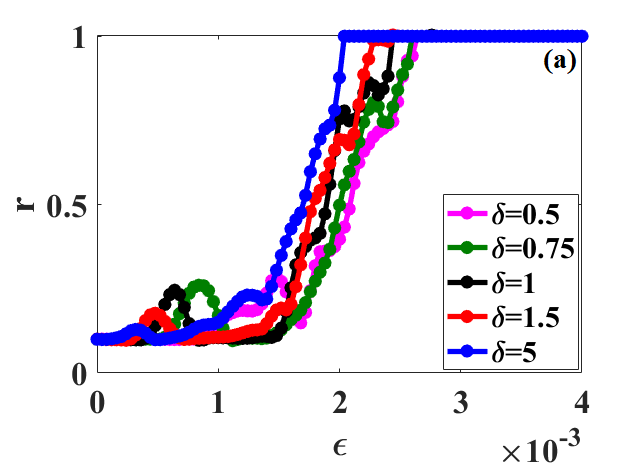}  &  
        \includegraphics[width=0.25\textwidth]{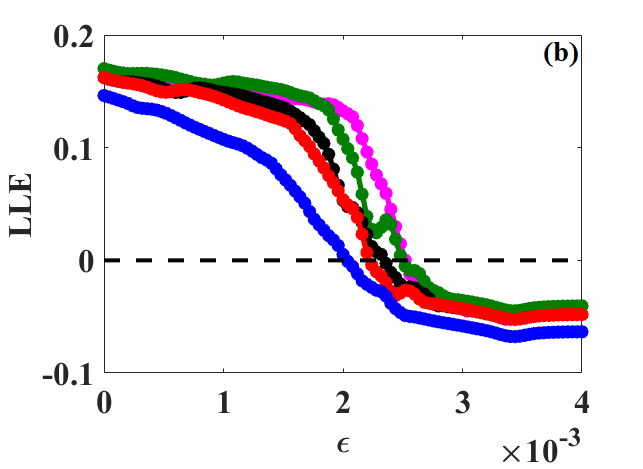} 
    \end{tabular}
    \caption{Transition to phase synchronization driven by dissipative energy. (a) Evolution of the order parameter $r$ and (b) Largest Lyapunov Exponent (LLE) as a function of the coupling strength $\epsilon$ for different values of the dissipative energy threshold $\delta_D$.}
\label{figdis}
\end{figure}

Fig.~\ref{figdis}(a) shows the influence of dissipative energy on the phase synchronization transition. The evolution of the order parameter $r$ is shown as a function of the coupling strength $\epsilon$ for several values of the dissipative energy threshold $\delta_D$. In contrast to the intrinsic-energy-based approach—where increasing $\delta$ leads to an first-order or explosive synchronization transition accompanied by hysteresis—the coupling mediated by dissipative energy does not promote such explosive behavior. Although the numerical framework remains identical (same parameter values $a=b=0.2$, $c=7$, same initial conditions, and consistent color coding for the thresholds), the synchronization transition modulated by dissipative energy remains smooth and continuous for all considered values of $\delta_D$. The order parameter increases progressively with the coupling strength $\epsilon$, indicating a second-order-like transition without any evidence of bistability or hysteresis. This qualitative difference is induced by the energy approach used to construct the network topology. In the intrinsic-energy-based approach, the threshold $\delta$ controls a structural selectivity: only oscillators with sufficiently close intrinsic energies can connect, which can generate strong correlations between network topology and dynamics and ultimately produce explosive synchronization. In contrast, dissipative energy reflects instantaneous energy exchange or losses occurring during interactions between oscillators. Because it is dynamically evolving rather than structurally fixed, the connectivity criterion defined by Eq.~\ref{eq::eq6} does not enforce persistent resonance-based constraints.  Instead, it leads to a more homogeneous and dynamically adaptive network organization, favoring a continuous emergence of synchronization.

Fig.~\ref{figdis}(b) illustrates the stability analysis of complete synchronization performed using the Master Stability Function (MSF) formalism in the case where the network topology is modulated by dissipative energy. In contrast to the intrinsic-energy-based case, where the transition of the Largest Lyapunov Exponent (LLE) exhibited an almost discontinuous behavior consistent with explosive synchronization, the LLE obtained here for dissipative-energy-driven topology shows a smooth and progressive transition from positive to negative values as the coupling strength increases. This progressive evolution indicates that the stabilization of the synchronization manifold occurs gradually, without abrupt structural–dynamical feedback. The behavior of the LLE is fully consistent with the evolution of the order parameter, confirming that dissipative-energy-based coupling leads to a second-order transition toward complete synchronization. 

\begin{figure}[htp!]
    \centering
    \begin{tabular}{c}
    \includegraphics[width=0.47\textwidth]{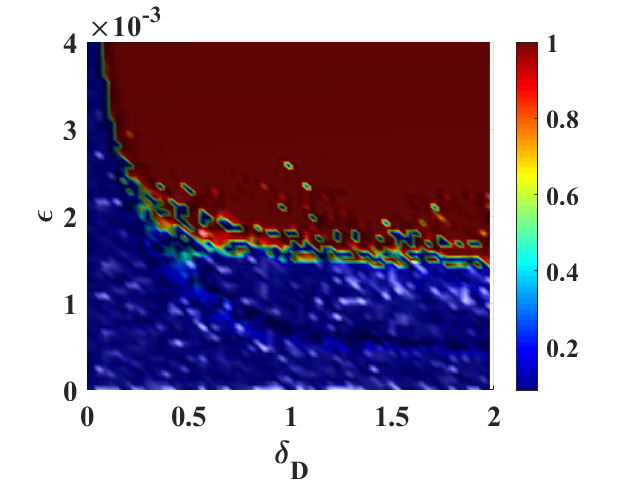}
    \end{tabular}
    \caption{Two-dimensional representation of the order parameter in the $(\delta_D,\epsilon)$ plane, showing the  synchronization transition. }
    \label{fig2dhdmin}
\end{figure}

For a more global analysis of synchronization in the framework of dissipative-energy-based approach, Fig.~\ref{fig2dhdmin} presents the evolution of the order parameter $r$ under a double variation of the dissipative connectivity threshold $\delta_D$ and the coupling strength $\epsilon$. The colorbar has the same meaning as in Fig.~\ref{fig2dhmin}, representing the magnitude of the order parameter $r$, which quantifies the level of phase synchronization.
The diagram clearly indicates two main regions in the $(\delta_D,\epsilon)$ parameter space: a blue region corresponding to low values of $r$ (desynchronized regime) and a red region where $r \to 1$, indicating global phase synchronization. The transition between these two regions defines the synchronization boundary. Unlike the intrinsic-energy-based approach, the transition line between desynchronization and synchronization in this case exhibits a smooth and continuous profile across the entire range of $\delta_D$, with only minor fluctuations as $\delta_D$ increases. This clearly illustrates that dissipative energy acts as a mechanism that inhibits explosive synchronization: it prevents the system from rapidly reaching a globally synchronized state, transforming what could be an abrupt, first-order-like transition into a smoother, second-order-like one.

\section{Effect of the system's state on the transition to phase synchronization}\label{sec:sec4}
This section is devoted to investigating how the internal dynamics of oscillators influence the onset of synchronization in a networked system. Here, ``internal dynamics'' refers to the qualitative nature of the isolated oscillator trajectories, which may exhibit either periodic or chaotic behavior depending on system parameters. In particular, we investigate how the level of periodicity or chaoticity of the R\"ossler oscillators affects the collective synchronization transition. The coupling mechanism between oscillators is the same as that described in Sec.~\ref{sec:MDM}, and is implemented in two distinct ways: through the conservative (intrinsic) energy term (see Eq.\ref{eq::eq5}) and through the dissipative energy term (see Eq.\ref{eq::eq6}). In both cases, the instantaneous network structure—due to the network’s temporal variability—is dynamically determined by the local energy values, allowing the connectivity to evolve over time. To investigate the influence of the internal dynamics, we select three representative values of the R\"ossler parameter $c$, while keeping $a=b=0.2$ fixed:
\begin{equation*}
    c \in \{4,\, 5.7,\, 8\}.
\end{equation*}
According to Fig.~\ref{fig1}, these values correspond to qualitatively distinct dynamical regimes of the isolated oscillator.
(i) For $c=4$, the system exhibits a multiperiodic regime characterized by approximately two dominant periods. (ii) For $c=5.7$, it operates in a well-known chaotic regime, with an infinite set of unstable periodic orbits embedded within a strange attractor. (iii) For $c=8$, the dynamics return to a multiperiodic regime, this time displaying approximately five distinct dominant periods (see Fig.~\ref{fig1} for the corresponding bifurcation diagram and maximum Lyapunov exponent). This choice of the values of the parameter $c$ allows for a systematic investigation of how qualitatively distinct dynamical regimes of the individual oscillators affect the onset and nature of the synchronization transition.

For each of these dynamical regimes, the effect on the network structure is analyzed by selecting three values of the parameters $\delta$ and $\delta_D$, which modulate the network topology based on the intrinsic and dissipative energy, respectively:
\begin{equation*}
    \delta,\, \delta_D \in \{0.5,\, 1,\, 2\}.
\end{equation*}
As previously described, these thresholds determine the existence of a connection between two oscillators at a given time: a link is established only if the absolute difference in the corresponding energy (conservative or dissipative) is below the threshold. By systematically exploring various values of both the internal dynamics parameter $c$ and the energy thresholds $\delta$ and $\delta_D$, the interaction between the oscillators’ microscopic dynamical regimes and the mesoscopic link formation process can be assessed, revealing how this interplay shapes the macroscopic synchronization transition.

\begin{figure*}[htp!]
\centering
\begin{tikzpicture}
\node[anchor=south west, inner sep=0] (tab) at (0,0) {
\begin{tabular}{cccc}
\raisebox{2.10 cm}{\boldsymbol{$\delta$}}  &
\includegraphics[width=0.30\textwidth]{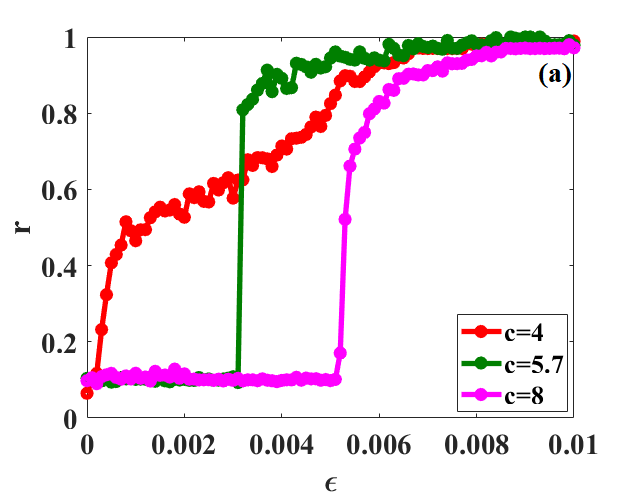} &
\includegraphics[width=0.30\textwidth]{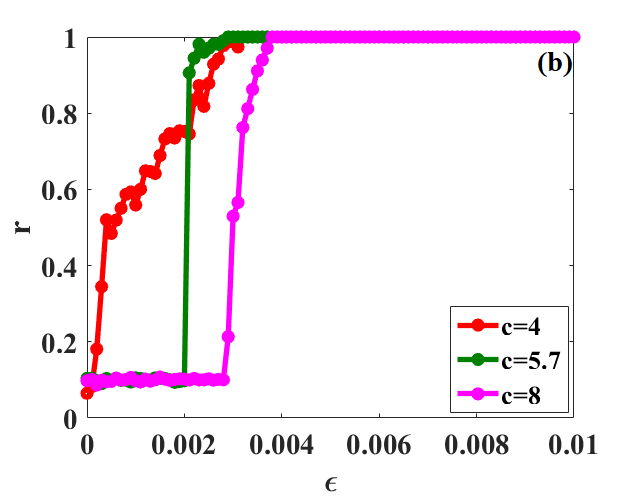} &
\includegraphics[width=0.30\textwidth]{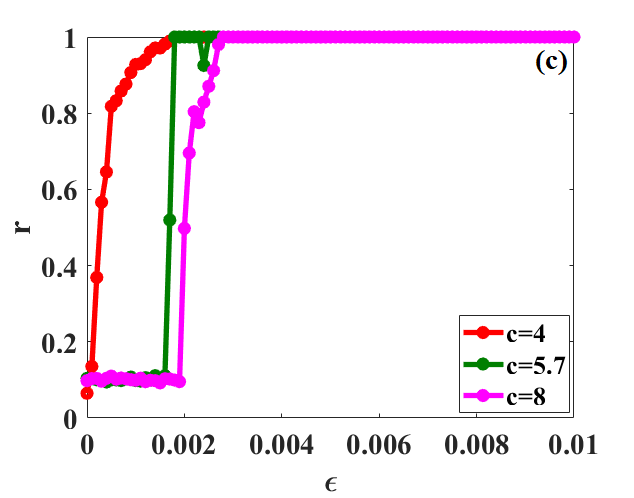} \\
\raisebox{2.2 cm}{\boldsymbol{$\delta_D$}} &
\includegraphics[width=0.30\textwidth]{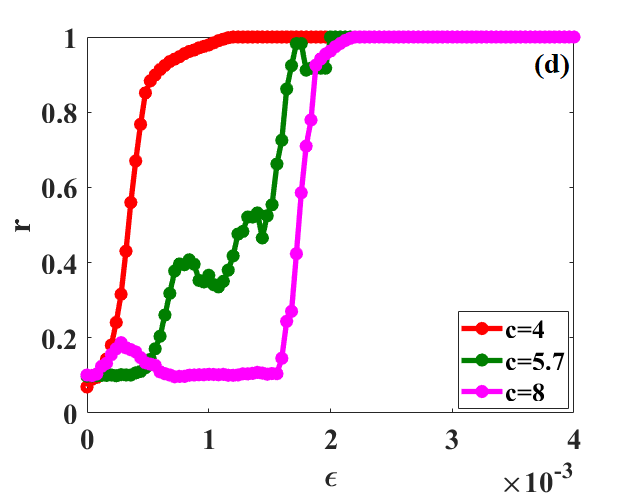} &
\includegraphics[width=0.30\textwidth]{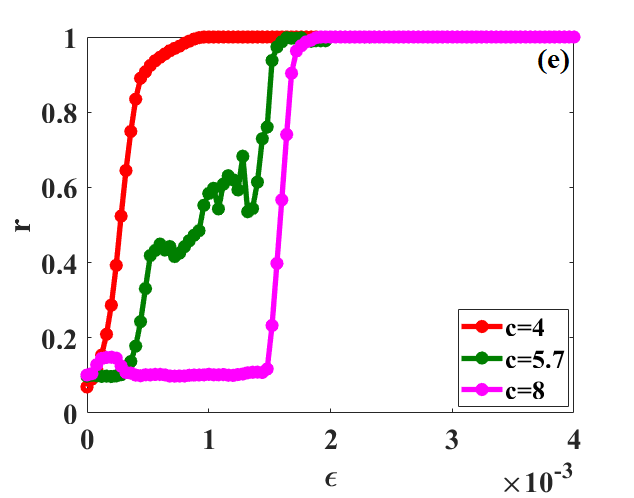} &
\includegraphics[width=0.30\textwidth]{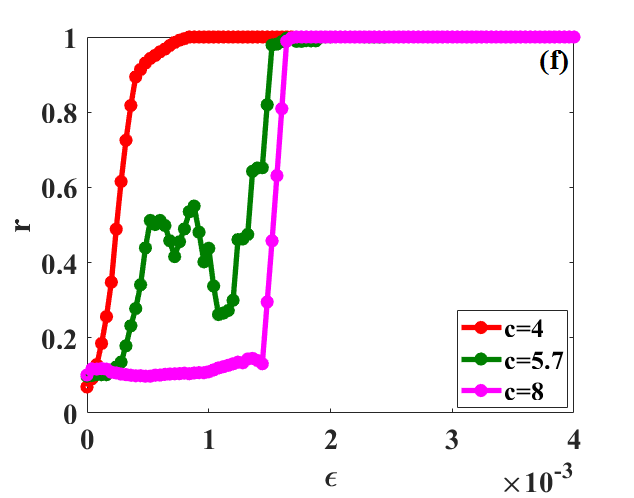} \\
\end{tabular}
};
\begin{scope}[x={(tab.south east)}, y={(tab.north west)}]
\node[anchor=south, yshift=-20pt] at (0.20,1.05)
{$\bm{\delta, \delta_D = 0.5}$};
\node[anchor=south, yshift=-20pt] at (0.52,1.05)
{$\bm{\delta, \delta_D = 1}$};
\node[anchor=south, yshift=-20pt] at (0.85,1.05)
{$\bm{\delta, \delta_D = 2}$};
\end{scope}
\end{tikzpicture}
\caption{Influence of internal dynamics on the synchronization transition.
The order parameter is plotted as a function of the coupling strength $\epsilon$ for varying values of $c$ and the threshold $\delta$ and $\delta_D$. Panels (a--c) show intrinsic-energy-driven dynamics, while panels (d--f) correspond to dissipative-energy-driven dynamics. Columns from left to right represent $\delta, \delta_D = 0.5$ (a,d), $\delta, \delta_D = 1$ (b,e), and $\delta, \delta_D = 2$ (c,f).}
\label{fig5}
\end{figure*}

Fig.~\ref{fig5} illustrates the effect of the internal dynamics of the oscillators on the transition to synchronization, as shown by the order parameter $r$ plotted as a function of the coupling strength $\epsilon$ for different dynamical regimes of the R\"ossler system, corresponding to $c=4, 5.7,$ and $8$. The results are shown for three values of the threshold $\delta$ and $\delta_D$, which modulate the instantaneous network connectivity based on differences in the intrinsic or dissipative energy. The first row (Fig.~\ref{fig5}(a--c)) corresponds to intrinsic-energy-driven modulation and demonstrates that the nature of the synchronization transition is strongly governed by the internal dynamical regime of the oscillators.
The coupling mediated by the intrinsic energy reveals a clear distinction in the nature of the synchronization transition depending on the local dynamics. For multiperiodic regimes, represented by $c=4$ and $c=8$, the transition gradually loses its explosive character and instead manifests as a smoother, continuous (second-order) transition, even as the network connectivity increases with larger $\delta$. In contrast, the chaotic regime ($c=5.7$) consistently maintains an explosive synchronization transition, as evidenced by the abrupt jumps in the green curves. This indicates that the level of internal complexity plays a critical role in promoting or suppressing the abrupt onset of synchronization. However, when the network connectivity is instead governed by the dissipative energy—shown in the second row of Fig.~\ref{fig5}(d--f)—explosive transitions are absent across all regimes, including both multiperiodic and chaotic cases. This suggests that dissipative coupling fundamentally alters the synchronization transition suppressing the abrupt transitions characteristic of intrinsic-energy-driven interactions. Overall, these results highlight that both the internal dynamics of the oscillators and the mechanism controlling the network topology (intrinsic versus dissipative energy) jointly determine whether the system exhibits explosive or continuous transition to synchronization.

\section{Extension to Lorenz model} \label{sec::LO}
This section aims to extend the analysis previously performed on the R\"ossler system to the Lorenz system, thereby demonstrating that the observed phenomena can be generalized to any dynamical system for which the Hamiltonian can be computed.

\subsection{Dynamics of an isolated Lorenz oscillator}

To generalize our study beyond the R\"ossler model, we consider the Lorenz system \cite{Lorenz1963}, which is a paradigmatic example of a chaotic dynamical system. It is governed by the set of these three ordinary differential equations given by Eq.~\ref{eq:lorenzlo}:
\begin{equation}
\label{eq:lorenzlo}
\begin{cases}
\dot{x} &= \sigma (y - x), \\
\dot{y} &= \rho x - y - x z, \\
\dot{z} &= x y - \beta z,
\end{cases}
\end{equation}
where $\sigma = 10$, $\rho = 28$, and $\beta = \dfrac{8}{3}$ are strictly positive parameters for which the system exhibits classical chaotic behavior. For these values, the system operates in a complex dynamical regime characterized by strong sensitivity to initial conditions. More generally, the system dynamics strongly depend on the parameter values and may display a wide variety of regimes, including periodic, multiperiodic, and chaotic behaviors.
\begin{figure*}[htp!]
    \centering
    \begin{tabular}{ccc}
    \includegraphics[width=0.33\textwidth]{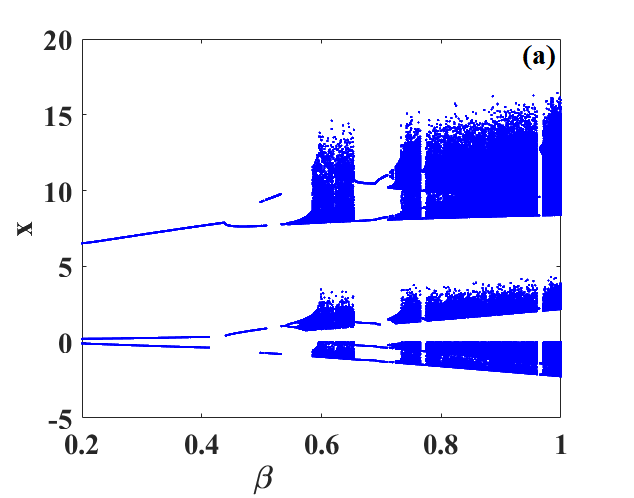}&
    \includegraphics[width=0.33\textwidth]{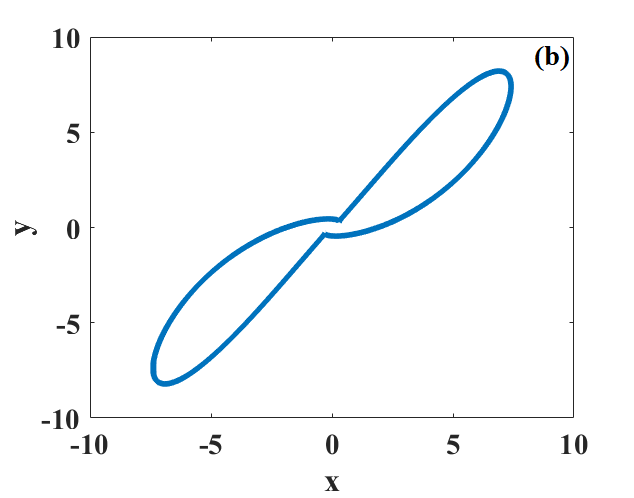}&        \includegraphics[width=0.33\textwidth]{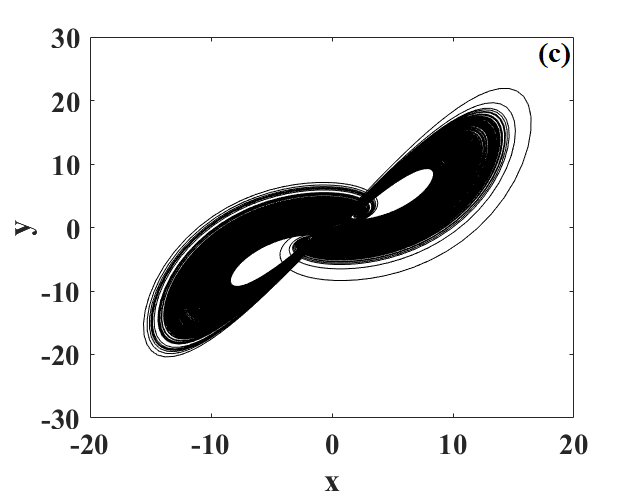}
    \end{tabular}      
   \caption{Dynamics of the Lorenz system with $\sigma = 10$ and $\rho = 28$: (a) shows the bifurcation diagram, while (b) and (c) display the corresponding attractors in the periodic ($\beta = 0.35$) and chaotic ($\beta = 1$) regimes, respectively.}
    \label{fig::figdynamiclo}
\end{figure*}
To illustrate this dynamical diversity, Fig.~\ref{fig::figdynamiclo}(a) shows the bifurcation diagram obtained by considering $\beta$ as the control parameter, varying within the interval $[0.2,\,1]$. This figure highlights an alternation of dynamical regimes. Periodic windows can be observed, as illustrated by the phase portrait shown in Fig.~\ref{fig::figdynamiclo}(b) for $\beta = 0.35$, where the trajectory converges to a stable periodic orbit. As $\beta$ increases, the system passes through multiperiodic regimes before reaching chaotic behavior, as shown in Fig.~\ref{fig::figdynamiclo}(c) for $\beta = 1$. In this case, the attractor exhibits a complex and aperiodic structure, which is a hallmark of chaotic dynamics.

To extend the classical Lorenz system defined by Eq.~\ref{eq:lorenzlo} to a networked setting, we consider a set of $N$ coupled Lorenz oscillators. 
\begin{equation}
\label{eq:lor}
\begin{cases}
\dot{x_j} &= \sigma (y_j - x_j) +\epsilon \sum_{i=1}^N G_{ji}(t) \left(x_i - x_j\right), \\
\dot{y_j} &= \rho x_j - y_j - x_j z_j, \\
\dot{z_j} &= x_j y_j - \beta z_j.
\end{cases}
\end{equation}
In this formulation, each oscillator/node $j$ with $j = 1,2,3,\ldots,N$ and $N = 100$, defined in Eq.~\ref{eq:lor} evolves according to the local Lorenz dynamics, while the coupling term $\epsilon \sum_{i=1}^N G_{ji}(t) (x_i - x_j)$ accounts for interactions with other nodes in the network. Here, $G_{ji}(t)$ denotes the time-dependent adjacency matrix of the network, which can be defined based on the intrinsic energy (see Eq.~\ref{eq::eq5}) and the dissipative energy (see Eq.~\ref{eq::eq6}); these energy definitions are provided in Appendix~\ref{Apen2}. The parameter $\epsilon$ represents the coupling strength between nodes.

\subsection{Interactions driven by intrinsic energy}
Following the same approach adopted in Sec.~\ref{sec::secIE} for the R\"ossler system with intrinsic energy, we extend the analysis to the model described in Eq.~\ref{eq:lor}. The same numerical integration procedure as that used for the R\"ossler system is implemented in order to ensure consistency between the two analyses.
\begin{figure}[htp!]
    \centering
    \begin{tabular}{ccc}
    \includegraphics[width=0.25\textwidth]{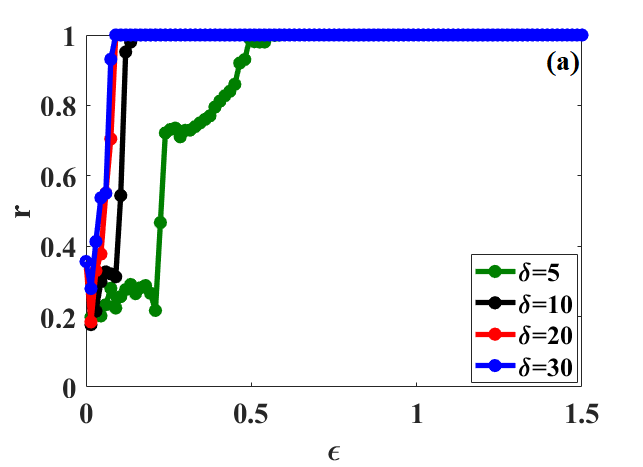}&
    \includegraphics[width=0.25\textwidth]{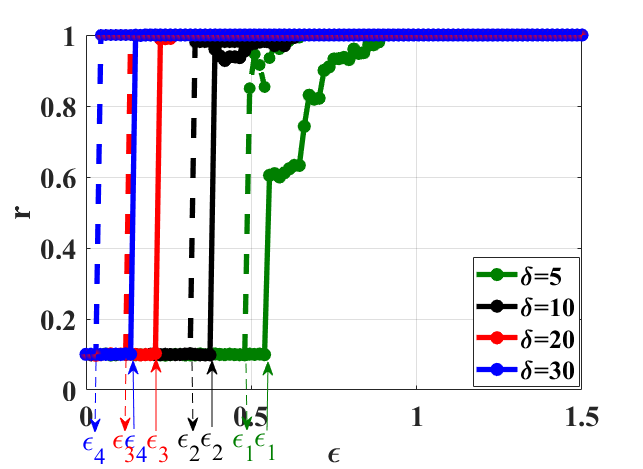}\\ 
    \end{tabular}      
    \caption{Synchronization transition in a network of coupled Lorenz systems, where the network topology is modulated by the intrinsic energy. The order parameter is shown as a function of the coupling strength $\epsilon$ for different values of the energy threshold $\delta$. Panel (a) corresponds to the periodic regime ($\beta = 0.35$), while panel (b) corresponds to the chaotic regime ($\beta = 1$). Solid curves indicate the forward process (increasing $\epsilon$), whereas dashed curves indicate the backward process (decreasing $\epsilon$), illustrating hysteresis.}
    \label{figlrz1}
\end{figure}
Therefore, Fig.~\ref{figlrz1} illustrates the evolution of the order parameter $r$ as a function of the coupling strength $\epsilon$ for a network of $N=100$ coupled Lorenz oscillators, considering four values of the intrinsic energy threshold
\begin{equation*}
\delta \in \{5, 10, 20, 30\}.
\end{equation*}
These values are selected according to the typical amplitude of the energy of the Lorenz system, as reported in Ref.~\cite{sarasola2004energy}. Fig.~\ref{figlrz1}(a) shows the synchronization transition when the local oscillators operate in a periodic regime with $\beta = 0.35$, whereas Fig.~\ref{figlrz1}(b) corresponds to the chaotic regime with $\beta = 1$. All other system parameters are kept constant.
In the periodic regime (see Fig.~\ref{figlrz1}(a)), the synchronization transition occurs gradually as the coupling strength $\epsilon$ increases. The order parameter increases continuously from the incoherent state toward full synchronization. Therefore, the transition remains smooth and does not exhibit any sharp discontinuity, indicating the absence of genuine explosive synchronization and instead giving rise to a second-order phase transition. This result is consistent with those obtained for the R\"ossler system.
In contrast, the chaotic regime (see Fig.~\ref{figlrz1}(b)) exhibits a pronounced and abrupt jump of the order parameter from $r \to 0$ to $r \to 1$. This discontinuous transition constitutes a clear signature of explosive synchronization, as confirmed by the hysteresis illustrated in Fig.~\ref{figlrz1}(b). Furthermore, increasing the intrinsic energy threshold $\delta$ results in a more densely connected network and accelerates the onset of explosive synchronization by reducing the critical coupling strength required for synchronization.

These observations demonstrate that chaos is a key ingredient for enabling explosive synchronization when the coupling is modulated by intrinsic energy. Importantly, the behavior observed here is in agreement with previous results obtained for networks of Rössler oscillators, confirming that intrinsic-energy–driven explosive synchronization is a robust phenomenon across different classes of chaotic systems.

\subsection{Interactions driven by the dissipative energy}

\begin{figure}[htp!]
    \centering
    \begin{tabular}{cc}       
    \includegraphics[width=0.25\textwidth]{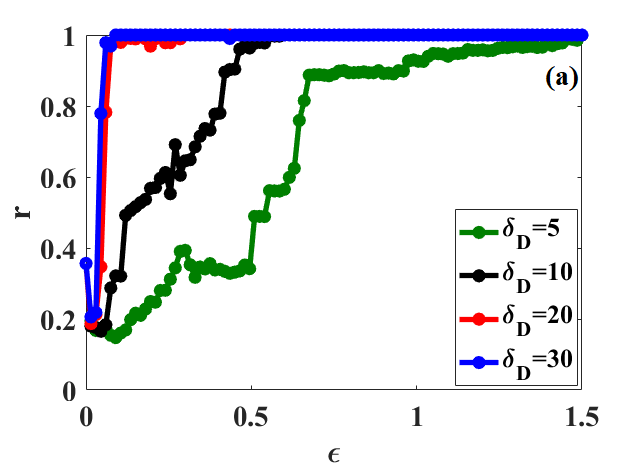}&
    \includegraphics[width=0.25\textwidth]{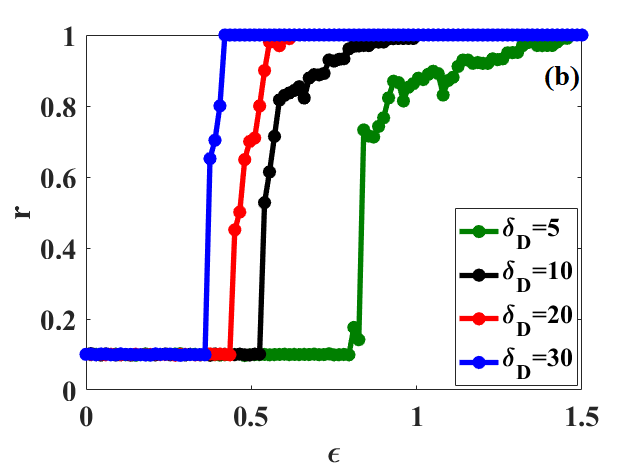}\\
    \end{tabular}      
    \caption{Influence of dissipative energy on the synchronization transition. The order parameter $r$ is plotted as a function of the coupling strength $\epsilon$ for four values of the threshold, $\delta_D \in \{5, 10, 20, 30\}$ modulating the network topology. Panel (a) corresponds to the periodic regime ($\beta = 0.35$), whereas panel (b) corresponds to the chaotic regime ($\beta = 1$).}
    \label{figlrz2}
\end{figure}

This subsection focuses on analyzing the impact of dissipative energy on the nature of the synchronization transition in the network of Lorenz oscillators described in Eq.~\ref{eq:lor}. Fig.~\ref{figlrz2} shows the evolution of the order parameter $r$ as a function of the coupling strength $\epsilon$ for four values of the dissipative energy threshold:
\begin{equation*}
    \delta_D \in \{5, 10, 20, 30\}.
\end{equation*}

In the periodic regime shown in Fig.~\ref{figlrz2}(a), where the local dynamics are in a periodic state, the transition toward synchronization develops gradually as the coupling strength increases. For all four values of $\delta_D$, the order parameter grows continuously, illustrating a second-order synchronization transition. We also note that increasing $\delta_D$ slightly reduces the critical coupling strength $\epsilon$ required to achieve synchronization. A similar second-order-like transition is observed when the local oscillator dynamics are in a chaotic state, as shown in Fig.~\ref{figlrz2}(b). These results indicate that dissipative energy acts as an inhibitory mechanism for collective organization, preventing the sudden coalescence of oscillators into a globally synchronized state. This behavior is fully consistent with previous findings for the R\"ossler network, confirming that dissipative interactions systematically suppress explosive synchronization across different classes of nonlinear oscillators.\\

\section{Conclusion} \label{sec: concl}

In this work, we introduced an energy-driven modeling framework to study synchronization 
dynamics in time-varying networks of chaotic and multiperiodic oscillators. Based 
on Hamiltonian 
formalism, we formulated a mechanism in which the network topology evolves according to 
energy similarities---either intrinsic or dissipative---between oscillators. Numerical simulations using the R\"ossler and Lorenz systems, with potential applicability to other dynamical models, demonstrate that this energy-based interaction scheme generates a wide range of synchronization transitions, encompassing both smooth continuous transitions---also called second-order transitions--- and abrupt explosive onsets---first order transitions. Our investigations show that the emergence of explosive synchronization critically depends on the interplay between the intrinsic dynamical regime of the oscillators and the energetic mechanism governing network connectivity. Chaotic dynamics combined with intrinsic-energy–driven interactions provide favorable conditions for explosive synchronization, whereas periodic and multiperiodic regimes systematically lead to smooth second-order transitions. In contrast, connectivity governed by dissipative energy suppress explosive synchronization and favour second-order synchronization transitions.
These results highlight the dual role of dynamics and topology in shaping collective behavior. They suggest that explosive synchronization is not solely a topological phenomenon, but emerges from a nontrivial interaction between the complexity of local dynamics and the rules governing adaptive connectivity. These findings reveal that explosive synchronization is not only a consequence of network topology, as commonly reported in studies based on paticular topologies such as scale-free or star networks, where abrupt transitions are commonly associated with structural heterogeneity or degree-frequency correlations \cite{chen2013explosive,gomez2011explosive,pinto2015explosive}. Instead, our results show that explosive transition may also emerge from an energy--dynamics--topology feedback, in which the local dynamical complexity of the oscillators and the energetic rule of link activation jointly determine the nature of the synchronization transition. However, the remaining question is now to understand how the energy-based interaction is related to the degree of the nodes and their local  distribution.
In future work, we aim to extend these findings to scenarios where the systems are non-identical or where the network exhibits heterogeneity in the local dynamics. Such extensions are motivated by the fact that real-world systems rarely consist of perfectly identical entities.

\section*{Acknowledgements}
T.N. acknowledges support from the “Reconstruction, Resilience and Recovery of Socio-Economic Networks” RECON-NET-EP\_FAIR\_005-PE0000013 “FAIR”-PNRR M4C2 Investment
1.3, financed by the European Union–NextGenerationEU. S.S.M acknowledges funding grant support from NBHM (National Board for Higher Mathematics), Department of Atomic Energy (DAE), Government of India for research project titled "Beyond Arnold Tongues: The Role of Extreme Curves, Superstable Curves, and Invariant Manifolds in the Bifurcation Structure of Chaotic Mapping", NBHM (R.P.)/ R\&D II/ 1028.

\section*{Data availability}
No data were created or analyzed in this study.

\appendix
\section{Network size effect on explosive synchronization}\label{Apen}

\begin{figure}[htp!]
    \centering
    \includegraphics[width=0.45\textwidth]{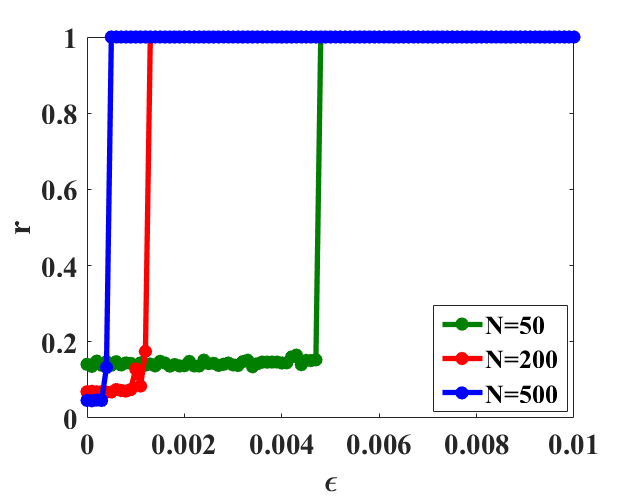}
    \caption{Evolution of the order parameter $r$ as a function of the coupling strength $\epsilon$ for different number of oscillators $N$, with $\delta = 5$.}
    \label{figrN}
\end{figure}

To demonstrate the independance of explosive synchronization illustrated in Sec.~\ref{sec::secIE} with respect to network size, Fig.~\ref{figrN} presents the order parameter $r$ defined in Eq.\ref{op} as a function of the coupling strength $\epsilon$ for different network sizes, namely $N = 50$, $200$, and $500$. In all three considered cases, the parameter $\delta = 5$, which modulates the network topology, is kept fixed. The system exhibits a sharp and discontinuous jump of the order parameter from a low-incoherence state (i.e., $r\to 0$) to a fully synchronized state (i.e., $r\to 1$), which is the hallmark of an explosive synchronization transition.
Although the critical coupling strength $\epsilon_c$ exhibits a slight shift as the network size $N$ increases, the abrupt nature of the transition remains preserved. The persistence of the discontinuous jump in $r$ for all investigated values of $N$ indicates that the phenomenon is not a finite-size artifact but an intrinsic dynamical property of the system.
Therefore, increasing the number of nodes does not modify the fundamental mechanism underlying the transition; it only influences the critical value of this explosive synchronization. These results provide strong evidence for the structural stability and robustness of explosive synchronization, supporting its persistence in the thermodynamic limit.

\section{Computation of the Master Stability Function}\label{Apen1}

To analyze the stability of the synchronized state, we highlight key aspects of the Master Stability Function (MSF) framework, which provides a sophisticated approach for determining the stability of synchronization in networks of coupled dynamical systems~\cite{pecoraPhysRevLett,njougouo2020,nguefoue2022network}.
Let us consider a network of $N$ identical coupled oscillators, where the dynamics of each isolated oscillator $j$ is governed by:
\begin{equation}
\label{Eqs1}
\mathbf{\dot{X}_j} = \mathbf{F}(\mathbf{X}_j),
\end{equation}
where $j=1,\cdots,N$, $\mathbf{X} \in \mathbb{\Re}^m$ denotes the state vector and $\mathbf{F} \in \mathbb{\Re}^m \to \mathbb{\Re}^m$ defines the local nonlinear dynamics.
By taking into account the temporal interactions between the oscillators in the network (see Eq.~\ref{eq::eq5} and Eq.~\ref{eq::eq6}), the dynamics of the $j^{\mathrm{th}}$ node can be expressed as
\begin{equation}
\mathbf{\dot{X}}_j = \mathbf{F}(\mathbf{X}_j) + \epsilon \sum_{i=1}^{N} \mathbf{L}_{ji}(t)\, \mathbf{W}(\mathbf{X}_i)
\label{eqs2}
\end{equation}
where $\epsilon$ represents the coupling strength, $\mathbf{W}:\mathbb{\Re}^m \rightarrow \mathbb{\Re}^m$ is an arbitrary coupling function of the node variables, and $\mathbf{L}(t)$ is the temporal Laplacian matrix of the network. The oscillators are considered to be synchronized when they evolve identically toward the same trajectory $\mathbf{s}(t)$, i.e.,
\begin{equation}
    \mathbf{X}_1 = \mathbf{X}_2 = \cdots = \mathbf{X}_N = \mathbf{s}.
    \label{eqs}
\end{equation}
For the sake of notational simplicity, we compact the local dynamics and the coupling functions of all network nodes into the following form:
\begin{equation}
 \mathbf{W}(\mathbf{X}) = [\mathbf{W}(\mathbf{X}_1),\, \mathbf{W}(\mathbf{X}_2),\, \ldots,\, \mathbf{W}(\mathbf{X}_N)],
\label{eqs4}
\end{equation}
in which, $\mathbf{W}(\mathbf{X})$ collects the coupling functions of all nodes.
\begin{equation}
\mathbf{F}(\mathbf{X}) = [\mathbf{F}(\mathbf{X}_1),\, \mathbf{F}(\mathbf{X}_2),\, \ldots,\, \mathbf{F}(\mathbf{X}_N)],
\label{eqs5}
\end{equation}
defined the vector of local dynamics of all nodes,
with
\begin{equation}
 \mathbf{X} = [\mathbf{X}_1,\, \mathbf{X}_2,\, \ldots,\, \mathbf{X}_N].
\label{eqs3}
\end{equation}
By combining Eqs.~\ref{eqs4}–\ref{eqs3}, the network dynamics can be compactly written as
\begin{equation}
\mathbf{\dot{X}} = \mathbf{F}(\mathbf{X}) + \epsilon\, \mathbf{L}(t) \otimes \mathbf{W}(\mathbf{X}),
\label{eqs6}
\end{equation}
where $\mathbf{L}(t)$ is the time-dependent Laplacian of the network and $\otimes$ denotes the Kronecker product.

Let us now define each term in Eq.~\ref{eqs6} within the framework of our analysis, considering the R\"ossler system as the local dynamics of each network node. For this particular case, the state of each node $j$ is represented by the vector $\mathbf{X}_j = (x_j, y_j, z_j)$, so that $\mathbf{F}(\mathbf{X}_j)$ and $\mathbf{W}(\mathbf{X}_j)$ are given by:
\begin{equation}
\mathbf{F}(\mathbf{X}_j)=
\begin{cases}
-\;y_j \;-\; z_j \\
x_j + a\,y_j \\
b + (x_j - c)z_j
\end{cases},
\quad\text{and}\quad
\mathbf{W}(\mathbf{X}_j) = 
\begin{pmatrix}
1 & 0 & 0 \\
0 & 0 & 0 \\
0 & 0 & 0
\end{pmatrix},
\label{eqs7}
\end{equation}
where $a = 0.2$, $b = 0.2$, and $c = 5.7$ are the system parameters. For these parameter values, the system exhibits chaotic behavior~\cite{ROSSLER1976397}.
We assume that all nodes in the network evolve synchronously, so that the equality given in Eq.~\ref{eqs} holds true at all times $t$. To analyze the stability of the synchronous solution $\mathbf{s}(t)$, we consider a small perturbation $\delta \mathbf{X}_j$ associated with node $j$. After the perturbation, the state of the $j^{\mathrm{th}}$ oscillator can be expressed as
\[
\mathbf{X}_j = \mathbf{s} + \delta \mathbf{X}_j,
\] 
which can be written compactly for the entire network as
\[
\delta \mathbf{X} = [\delta \mathbf{X}_1;\, \delta \mathbf{X}_2;\, \ldots;\, \delta \mathbf{X}_N].
\]

Substituting the perturbed state $\mathbf{X}_j = \mathbf{s} + \delta \mathbf{X}_j$ into Eq.~\ref{eqs2} and performing a first-order Taylor expansion of $\mathbf{F}(\mathbf{s} + \delta \mathbf{X}_j)$ and $\mathbf{H}(\mathbf{s} + \delta \mathbf{X}_j)$, we obtain the following variational equation:
\begin{equation}
\delta \mathbf{\dot{X}}_j = \mathbf{DF}(\mathbf{s})\, \delta \mathbf{X}_j 
+ \epsilon \sum_{i=1}^{N} \mathbf{L}_{ji}(t)\, \mathbf{DW}(\mathbf{s})\, \delta \mathbf{X}_i,
\label{eqs8}
\end{equation}
where $\mathbf{DF}(\mathbf{s})$ and $\mathbf{DW}(\mathbf{s})$ are the $m \times m$ Jacobian matrices of the corresponding vector functions $\mathbf{F}$ and $\mathbf{W}$ evaluated at the synchronous state $\mathbf{s}$. According to the tensor notation, Eq.~\ref{eqs8} can be rewritten in a more simple tensor-product form for the entire network as:
\begin{equation}
\delta \mathbf{\dot{X}} = \left[ \mathbf{I}_N \otimes \mathbf{DF}(\mathbf{s})
      + \epsilon\, \mathbf{L}(t) \otimes \mathbf{DW}(\mathbf{s}) \right] \delta \mathbf{X},
\label{eqs9}
\end{equation}
Given the form of Eq.~\ref{eqs9}, which is a linear system of ordinary differential equations describing the evolution of infinitesimal perturbations around the synchronous manifold. Its solutions can be formally written as
\[
\delta \mathbf{X}_j = \exp(\lambda_j t),
\]
where the exponent $\lambda$ characterizes the growth or decay of perturbations: the synchronous state is unstable if $\lambda > 0$ and stable if $\lambda < 0$.\\
However, in the tensor-product form, the perturbation components remain coupled through the network Laplacian $\mathbf{L}(t)$. In order to analyze the stability of the synchronized solution, it is convenient to decouple these equations by projecting the perturbation dynamics onto the eigenbasis of the Laplacian matrix at each time $t$.

Diagonalizing the coupling term in Eq.~\ref{eqs9} using the eigenvectors of $\mathbf{L}(t)$ therefore transforms the system into a set of independent variational equations associated with the Laplacian eigenmodes. For the $k$th mode, we obtain
\begin{equation}
\delta \mathbf{\dot{X}}_k =
\left[
\mathbf{DF}(\mathbf{s})
+
\epsilon\,\alpha_k(t)\,\mathbf{DW}(\mathbf{s})
\right]
\delta \mathbf{X}_k ,
\label{eqs10}
\end{equation}
where $\alpha_k$ ($k = 1,\ldots,N$) denote the eigenvalues of the Laplacian matrix. Introducing the composite parameter $\alpha = \epsilon \alpha_k$, the variational equation can be written in the universal form
\begin{equation}
\dot{\delta \mathbf{X}} =
\left[
\mathbf{DF}(\mathbf{s})
+
\alpha(t)\,\mathbf{DW}(\mathbf{s})
\right]
\delta \mathbf{X},
\label{eqs11}
\end{equation}
which is known as the \emph{Master Stability Equation} (MSF)\cite{pecoraPhysRevLett}. 
Because the local dynamics is chaotic, its Jacobian matrix depend on the state variables of the system and therefore implicitly on time. As a consequence, Eq.~\ref{eqs11} becomes a time-dependent linear system and an analytical determination of the MSF is generally not possible. The stability analysis is therefore performed numerically by computing the largest Lyapunov exponent (LLE) associated with the variational equation. The synchronous state is stable when the LLE is negative, which indicates that transverse perturbations decay exponentially. Conversely, synchronization becomes unstable when the LLE is positive, as perturbations grow with time.

\section{Calculation of the Energy in the Lorenz System\label{Apen2}}

Let us now outline the calculation of the intrinsic energy $\mathbf{H}_j$ and the dissipative energy rate $\dot{\mathbf{H}}_j$ associated with each node $j$ of the Lorenz system defined in Eq.~\ref{eq:lor}. To this end, we employ the Hamiltonian formalism described in Sec.~\ref{sec::HF}. Therefore, according to Helmoholtz's theorem, the dynamical field $\mathbf{F}(\mathbf{X}_j(t))$ of the system $j$ at time $t$ can be written as the sum
\[
\mathbf{F}(\mathbf{X}_j(t)) = \mathbf{F}_c(\mathbf{X}_j(t)) + \mathbf{F}_d(\mathbf{X}_j(t)).
\]
$\mathbf{F}_c(\mathbf{X}_j(t))$ is a divergence-free vector field representing the conservative (rotational) component of the dynamics, whereas $\mathbf{F}_d(\mathbf{X}_j(t))$ is a gradient vector field corresponding to the dissipative component and accounting for the total divergence of the system. $\mathbf{X}_j=(x_j,y_j,z_j)$ represents the state vector of the oscillator $j$.
Following this criterion, the dynamics of the $j$th Lorenz oscillator can be decomposed as
\begin{equation}
\begin{aligned}
\mathbf{F}_c(\mathbf{X}_j(t)) &= 
\begin{pmatrix}
\sigma y_j\\
\rho x_j - x_j z_j\\
x_j y_j
\end{pmatrix}, \text{and} \quad
\mathbf{F}_d(\mathbf{X}_j(t)) = 
\begin{pmatrix}
- \sigma x_j\\
- y_j\\
- \beta z_j
\end{pmatrix}.
\end{aligned}
\label{eq:FdFc}
\end{equation}
This decomposition allows the construction of a generalized intrinsic energy function $\mathbf{H}_j$ and the explicit computation of its time derivative $\mathbf{\dot H}_j$, thereby quantifying the energy dissipation at the level of each oscillator.\\
According to Eq.~\ref{eq::eq3}, the intrinsic energy $\mathbf{H}_j$ satisfies the following condition
\begin{equation}
\sigma y_j \frac{\partial \mathbf{H}_j}{\partial x_j} + (\rho x_j - x_j z_j) \frac{\partial \mathbf{H}_j}{\partial y_j} + x_j y_j \frac{\partial \mathbf{H}_j}{\partial z_j} = 0,
\end{equation}
where an analytical resolution leads to the following quadratic solution given by Eq.~\ref{eqlo}:
\begin{equation}
\mathbf{H}_j = \frac{1}{2} \Big(- \frac{\rho}{\sigma} x_j^2 + y_j^2 + z_j^2 \Big).
\label{eqlo}
\end{equation}
According to Eq.~\ref{eq::eq4}, the dissipative energy flow is then given by 
\begin{equation}
\mathbf{\dot H_j} = \rho x_j^2 - y_j^2 - \beta z_j^2,
\end{equation}
which quantifies the instantaneous energy dissipated or injected by each oscillator.


\end{document}